\documentclass[bibyear]{aa}
\usepackage{graphicx}
\usepackage{txfonts}
\usepackage{here}
\newcommand{\whya}{\mbox{W~Hya}}
\newcommand{\RSTAR}{\mbox{$R_{\star}$}}
\newcommand{\LSOL}{\mbox{$L_{\sun}$}}

\newcommand{\MSOL}{\mbox{$M_{\sun}$}}

\newcommand{\micron}{\mbox{$\mu$m}}
\newcommand{\KMS}{\mbox{km s$^{-1}$}}
\newcommand{\alumina}{\mbox{Al$_2$O$_3$}}
\newcommand{\corundum}{\mbox{Al$_2$O$_3$}}
\newcommand{\forsterite}{\mbox{Mg$_2$SiO$_4$}}
\newcommand{\enstatite}{\mbox{MgSiO$_3$}}
\newcommand{\Ha}{\mbox{H$\alpha$}}
\newcommand{\HOH}{\mbox{H$_2$O}}
\newcommand{\wlc}{\mbox{$\lambda_c$}}

\newcommand{\Qplus}{\mbox{$Q_+$}}
\newcommand{\Qminus}{\mbox{$Q_-$}}
\newcommand{\Uplus}{\mbox{$U_+$}}
\newcommand{\Uminus}{\mbox{$U_-$}}
\newcommand{\IP}{\mbox{$I_{\rm P}$}}
\newcommand{\PL}{\mbox{$p_{\rm L}$}}
\newcommand{\alfcenA}{\mbox{$\alpha$~Cen~A}}
\newcommand{\tauV}{\mbox{$\tau_{500{\rm nm}}$}}
\usepackage{color}

\begin{document}

\title{
Clumpy dust clouds and extended atmosphere of the AGB star W~Hya 
revealed with VLT/SPHERE-ZIMPOL and VLTI/AMBER II
\thanks{
Based on SPHERE and AMBER observations made with the Very Large Telescope 
and Very Large Telescope Interferometer of the European Southern Observatory. 
Program ID: 095.D-0397(B) and 092.D-0461(A). 
}
}
\subtitle{
Time variations between pre-maximum and minimum light
}

\author{K.~Ohnaka\inst{1} 
\and
G.~Weigelt\inst{2} 
\and
K.-H.~Hofmann\inst{2} 
}

\offprints{K.~Ohnaka}

\institute{
Universidad Cat\'{o}lica del Norte, Instituto de Astronom\'{i}a, 
Avenida Angamos 0610, Antofagasta, Chile\\
\email{k1.ohnaka@gmail.com}
\and
Max-Planck-Institut f\"{u}r Radioastronomie, 
Auf dem H\"{u}gel 69, 53121 Bonn, Germany
}

\date{Received / Accepted }

\abstract
% Context
{}
% Aim
{
Our recent visible polarimetric images of the well-studied AGB 
star \whya\ taken at pre-maximum light (phase 0.92) with 
VLT/SPHERE-ZIMPOL have revealed clumpy dust clouds 
close to the star at $\sim$2~\RSTAR. 
We present second-epoch SPHERE-ZIMPOL observations of \whya\ 
at minimum light (phase 0.54) 
as well as high-spectral resolution long-baseline interferometric 
observations with the AMBER instrument at the Very Large Telescope 
Interferometer (VLTI). 
}
% Methods
{
We observed \whya\ with VLT/SPHERE-ZIMPOL at three wavelengths in the 
continuum (645, 748, and 820~nm), in the H$\alpha$ line at 656.3~nm, and 
in the TiO band at 717~nm.  
The VLTI/AMBER observations were carried out in the wavelength region 
of the CO first overtone lines near 2.3~\micron\ with a spectral 
resolution of 12\,000. 
}
% Results
{
The high-spatial resolution polarimetric images obtained with SPHERE-ZIMPOL 
have allowed us to detect clear time variations in the clumpy dust clouds 
as close as 34--50~mas (1.4--2.0~\RSTAR) to the star. 
We detected the formation of a new dust cloud as well as the disappearance of 
one of the dust clouds detected at the first epoch. 
The \Ha\ and TiO emission extends to $\sim$150~mas ($\sim$6~\RSTAR), 
and the \Ha\ images obtained at two epochs reveal time variations. 
The degree of linear polarization measured at minimum light, which ranges 
from 13 to 18\%, is higher than that observed at pre-maximum light. 
The power-law-type limb-darkened disk fit to the AMBER data in the continuum 
results in a limb-darkened disk diameter of 
$49.1 \pm 1.5$~mas and a limb-darkening parameter of 
$1.16 \pm 0.49$, indicating that the atmosphere is more extended with 
weaker limb-darkening compared to pre-maximum light. 
Our Monte Carlo radiative transfer modeling shows that the second-epoch 
SPHERE-ZIMPOL data can be explained by a shell of 0.1~\micron\ grains of 
\corundum, \forsterite, and \enstatite\ with a 550~nm optical depth of 
$0.6 \pm 0.2$ and an inner and outer radii of 1.3~\RSTAR\ and 
$10 \pm 2$~\RSTAR, respectively.  
Our modeling suggests the predominance of 
small (0.1~\micron) grains at minimum light, in marked contrast to the 
predominance of large (0.5~\micron) grains at pre-maximum light. 
}
% Conclusions
{
The variability phase dependence of the characteristic grain size 
implies that small grains might just have started to form at minimum light 
in the wake of a shock, 
while the pre-maximum light phase might have corresponded to the phase 
of efficient grain growth. 
}

\keywords{
techniques: polarimetric -- 
techniques: interferometric -- 
stars: imaging -- 
stars: AGB and post-AGB -- 
(stars:) circumstellar matter --
stars: individual: W~Hya
}   %  END OF ABSTRACT

\titlerunning{Clumpy dust formation and molecular outer atmosphere 
of the AGB star W~Hya II. Time variations}
\authorrunning{Ohnaka et al.}
\maketitle

\begin{table*}
\caption {
Summary of the SPHERE-ZIMPOL observations of \whya. 
}
\begin{center}

\begin{tabular}{l c c c c c c l l l l l}\hline
\# & $t_{\rm obs}$ & DIT  & NDIT & N$_{\rm exp}$ & N$_{\rm pol}$ 
& N$_{\rm dither}$ & Filter & Seeing & AM & Strehl & Strehl \\
   & (UTC)       & (sec)&      &               &      &        & (cam1/cam2) 
& (\arcsec) &  & ($H$) & (visible) \\
\hline
\multicolumn{12}{c}{\whya : 2016 March 23 (UTC)}\\
\hline
1 & 03:58:46 & 20& 6  & 1 & 1 & 3 & CntHa/NHa     & 0.73 & 1.16 & 0.86 & 0.38 \\
2 & 04:29:04 & 3 & 4  & 1 & 8 & 3 & TiO717/Cnt748 & 0.74 & 1.09 & 0.86 & 0.46 \\
3 & 05:05:18 & 1 & 12 & 1 & 6 & 3 & Cnt820/Cnt820 & 0.77 & 1.04 & 0.85 & 0.52 \\
\hline
\multicolumn{12}{c}{HIP67834: 2016 March 23 (UTC)}\\
\hline
C1 & 07:10:37 & 10 & 10 & 1 & 1 & 3 &  CntHa/NHa &    0.63 & 1.02 & 0.74 & 0.25\\
C2 & 07:35:09 &  5 & 2 & 1 & 4 & 3 &  TiO717/Cnt748 & 0.54 & 1.04 & 0.77 & 0.36\\
C3 & 07:49:30 &  2 & 6 & 1 & 5 & 3 &  Cnt820/Cnt820 & 0.69 & 1.07 & 0.73 & 0.38\\
\hline
\label{obs_log_sphere}
\vspace*{-7mm}

\end{tabular}
\end{center}
\tablefoot{
DIT: Detector integration time.  
NDIT: Number of frames.
N$_{\rm exp}$: Number of exposures of each polarization component in each 
polarization cycle at each dithering position. 
N$_{\rm pol}$: Number of polarization cycles at each dithering position. 
N$_{\rm dither}$: Number of dithering positions. 
Seeing was measured in the visible. 
AM: Airmass. 
The Strehl ratios in the visible are computed from the $H$-band Strehl ratios 
for \whya, while they were measured from the observed ZIMPOL images for 
HIP67834. 
}
\end{table*}

\section{Introduction}
\label{sect_intro}

Improving our understanding of dust formation is indispensable for 
solving the long-standing problem of the mass loss from 
asymptotic giant branch (AGB) stars.  
It is considered that the dust formation takes place within a few stellar 
radii, and observational studies of this region is crucial for 
clarifying the wind acceleration mechanism in AGB stars. 
The angular size of this innermost key region is smaller than $\sim$100~mas 
even for nearby AGB stars, which means that we need high spatial resolution 
to study the dust formation zone. 

The star \whya\ is one of the closest AGB stars with 
a distance of $78^{+6.5}_{-5.6}$~pc (Knapp et al. \cite{knapp03}), and 
therefore, it has been well studied from the visible to the radio. 
In the Simbad database, W~Hya is classified as a semi-regular variable. 
However, while its variability amplitude ($\Delta V \! \approx \! 3$) is much 
smaller than that of typical Mira stars, clear periodicity can be seen in its 
light curve, and therefore, it is often classified as Mira in the literature. 
Recently we carried out polarimetric imaging of \whya\ 
at five wavelengths from 645 to 820 nm with the VLT instrument 
SPHERE-ZIMPOL (Ohnaka et al. \cite{ohnaka16}, hereafter Paper~I).  
Taking advantage of the superb performance of extreme adaptive optics (AO), 
we succeeded in spatially resolving 
clumpy dust clouds very close to the star at $\sim$50 mas 
(just $\sim$2 \RSTAR). 
Our 2-D radiative transfer modeling suggests the predominance of large 
(0.4--0.5~\micron) grains of corundum (\corundum) or iron-poor silicate 
such as \forsterite\ and \enstatite\ and 
a density enhancement of a factor of $4\pm 1$ in the clumps.  
Our results are consistent with the near-IR polarimetric interferometric 
observations obtained with the aperture-masking technique (Norris et al. 
\cite{norris12}), which suggested the presence of grains of 0.3~\micron\ 
at $\sim$2~\RSTAR. 
Grains such as \corundum, \forsterite, and \enstatite\ do not efficiently 
absorb stellar photons in the visible and in the near-IR due to their low 
opacities at these wavelengths.  This means that the dust temperature can 
remain lower than the sublimation temperature, and therefore, 
the dust grains can form very close to the star.  
The detection of large, transparent grains close to the star 
lends support to the scenario proposed by H\"ofner (\cite{hoefner08}) 
that the radiation pressure 
due to the scattering of stellar photons---instead of absorption---by 
large, micron-sized grains can drive the mass loss in oxygen-rich AGB stars. 

Dust formation in Mira-type AGB stars is strongly affected by the periodic 
passage of shocks generated by large-amplitude stellar pulsation, 
as demonstrated by self-consistent hydrodynamical models for AGB stellar winds 
(e.g., Fleischer et al. \cite{fleischer91}, \cite{fleischer92} for 
carbon-rich stars, Jeong et al. \cite{jeong03} and H\"ofner et al. 
\cite{hoefner16} for oxygen-rich stars). 
The nucleation and growth process of dust grains from the gas 
is a complicated, non-equilibrium process, which is not yet completely 
understood (e.g., Gobrecht et al. \cite{gobrecht16}; 
Gail et al. \cite{gail16}). 
Therefore, 
studying time variations in the physical properties of the dust formation 
region is crucial for understanding the grain nucleation and growth process. 
The clear periodicity in the light curve of \whya\ suggests the presence 
of pulsation, although its amplitude may be smaller than that of typical 
Miras. This means that we can expect the pulsation to cause shocks in 
the atmosphere of \whya, which affect the formation of dust clouds.

In the present paper, we report on second-epoch polarimetric imaging 
observations of \whya\ with VLT/SPHERE-ZIMPOL as well as  
high-spectral resolution long-baseline interferometric 
observations in the CO first overtone lines near 2.3~\micron\ 
with VLTI/AMBER to study time variations in the gas and dust environment 
in the immediate vicinity of the star. 
In Sect.~\ref{sect_obs}, we describe the observations and data reduction. 
The observational results are presented in Sect.~\ref{sect_res}. 
The determination of the effective temperature and luminosity is 
described in Sect.~\ref{sect_stellar_param}.  The radiative transfer modeling 
of the data is presented in Sect.~\ref{sect_modeling}, followed by 
interpretation of the results in Sect.~\ref{sect_discuss} and concluding 
remarks in Sect.~\ref{sect_concl}.

\begin{table}
\begin{center}
\caption {
The flux of \whya\ derived at the second epoch (phase 0.54, minimum light) 
with five filters of our SPHERE-ZIMPOL observations. 
These values are used for the flux calibration of the SPHERE-ZIMPOL 
images.  The flux obtained at the first epoch (phase 0.92, pre-maximum 
light) from Paper~I is also given for reference. 
}

\begin{tabular}{l l c c}\hline
Filter & $\lambda_{\rm c}$ &  Flux (phase 0.54) & Flux (phase 0.92) \\
       & (nm)            &  (W~m$^{-2}$~$\micron^{-1}$) & (W~m$^{-2}$~$\micron^{-1}$) \\
\hline
CntHa  & 644.9  & $5.62\times10^{-11}$ & $3.78\times10^{-10}$ \\
NHa    & 656.34 & $6.40\times10^{-11}$ & $4.33\times10^{-10}$ \\
TiO717 & 716.8  & $2.22\times10^{-10}$ & $1.09\times10^{-9}$ \\
Cnt748 & 747.4  & $5.30\times10^{-10}$ & $4.36\times10^{-9}$ \\
Cnt820 & 817.3  & $5.29\times10^{-10}$ & $1.05\times10^{-8}$ \\
\hline
\label{whya_photometry}
\vspace*{-7mm}

\end{tabular}
\end{center}
\end{table}

\begin{table*}
\caption {
Summary of the VLTI/AMBER observations of \whya\ and the calibrator \alfcenA. 
}
\begin{center}

\begin{tabular}{l c c c c r l }\hline
\# & $t_{\rm obs}$ & $B_{\rm p}$ & PA     & Seeing   & $\tau_0$ &
${\rm DIT}\times{\rm N}_{\rm f}\times{\rm N}_{\rm exp}$ \\ 
& (UTC)       & (m)       & (\degr) & (\arcsec)       &  (ms)    &  (ms)  \\
\hline
\multicolumn{7}{c}{\whya: 2014 February 11 (UTC)}\\
\hline
1 & 08:23:25 & 11.30/22.63/33.94 & 21/21/21    & 1.06 & 4.1 & $120\times500\times1$ \\
2 & 08:25:32 & 11.30/22.64/33.94 & 22/22/22    & 1.06 & 4.1 & $120\times500\times1$ \\
3 & 08:48:33 & 11.30/22.62/33.92 & 24/24/24    & 0.81 & 5.3 & $120\times500\times1$ \\
4 & 08:50:40 & 11.30/22.62/33.92 & 25/25/25    & 0.94 & 4.6 & $120\times500\times1$ \\
5 & 08:52:48 & 11.29/22.62/33.91 & 25/25/25    & 1.09 & 3.9 & $120\times500\times1$ \\
6 & 08:54:56 & 11.29/22.61/33.91 & 25/25/25    & 0.98 & 4.4 & $120\times500\times1$ \\
7 & 08:57:02 & 11.29/22.61/33.90 & 25/25/25    & 0.79 & 5.3 & $120\times500\times1$ \\
\hline
\multicolumn{7}{c}{\alfcenA: 2014 February 11 (UTC)}\\
\hline
C1 & 05:02:03 & 10.07/20.17/30.23 & 165/165/165 & 1.02& 4.8 &$120\times500\times5$\\
C2 & 05:36:38 & 10.14/20.32/30.46 & 171/171/171 & 0.97& 5.0&$120\times500\times5$\\
C3 & 06:51:22 & 10.19/20.40/30.59 & 3/3/3 & 0.94& 4.9 &$120\times500\times5$\\
C4 & 07:26:01 & 10.15/20.33/30.48 & 9/9/9 & 0.82& 5.5 &$120\times500\times5$\\
C5 & 08:02:11 & 10.07/20.18/30.25 & 14/14/14 & 1.26& 3.5&$120\times500\times5$\\
C6 & 09:06:12 & 9.83/19.69/29.52 & 24/24/24 & 0.98& 4.2&$120\times500\times5$\\
C7 & 09:40:40 & 9.64/19.31/28.95 & 30/30/30 & 0.84& 3.8&$120\times500\times5$\\
\hline
\label{obs_log_amber}
\vspace*{-7mm}

\end{tabular}
\end{center}
\tablefoot{
$B_{\rm p}$: Projected baseline length.  PA: Position angle of the baseline 
vector projected onto the sky. 
DIT: Detector Integration Time.  $N_{\rm f}$: Number of frames in each 
exposure.  $N_{\rm exp}$: Number of exposures. 
The seeing and the coherence time ($\tau_0$) were measured in the visible. 
}
\end{table*}

\section{Observations}
\label{sect_obs}

\subsection{Visible polarimetric imaging observations with SPHERE-ZIMPOL}
\label{subsect_obs_zimpol}

VLT/SPHERE allows us to carry out high-spatial resolution and high-contrast 
imaging, taking advantage of an extreme adaptive optics (AO) system for the 
wavelength range from 0.55 to 2.32~\micron\ (Beuzit et al. \cite{beuzit08}).  
The ZIMPOL instrument is a unit for nearly diffraction-limited 
polarimetric imaging (also classical, non-polarimetric imaging) 
at 550--900~nm (Thalmann et al. \cite{thalmann08}).  
Our SPHERE-ZIMPOL observations of \whya\ 
(Program ID: 095.D-0397(B), P.I.: K.~Ohnaka) occurred 
on 2016 March 23 (UTC). 
As in our first-epoch observations reported in Paper~I, 
we used the P2 mode, in which the field orientation remains fixed.  
The $V$ magnitude of \whya\ at the time of our second-epoch 
observations is estimated to be $\sim$9.5, corresponding to phase 0.54 
(minimum light), from the light curve of the 
American Association of Variable Star Observers (AAVSO).  
We observed the K1III star HIP67834 ($V$ = 7.6) as a reference of the point 
spread function (PSF). 
Given an angular diameter of $0.538\pm 0.016$~mas 
(CalVin database\footnote{http://www.eso.org/observing/etc/bin/gen/\\
form?INS.NAME=CALVIN+INS.MODE=CFP}), this reference star should 
appear as a point source with the spatial resolution of SPHERE-ZIMPOL.  

As in our first-epoch observations, we used three filters to sample the 
\mbox{(pseudo-)}continuum regions relatively (but not entirely) free from 
the TiO bands (CntHa with central wavelength \wlc\ = 644.9~nm and FWHM = 4.1~nm, 
Cnt748 with \wlc\ = 747.4~nm and FWHM = 20.6~nm, and Cnt820 with 
\wlc\ = 817.3~nm and FWHM = 19.8~nm). 
In addition, we observed with the NHa filter (\wlc\ = 656.34~nm and 
FWHM = 0.97~nm) and the TiO717 filter (\wlc\ = 716.8~nm and FWHM = 19.7~nm) 
to probe the \Ha\ emission and TiO emission, respectively. 
The SPHERE-ZIMPOL instrument is equipped with two cameras (cam1 and cam2), 
with a pixel scale of 3.628~mas. 
This allows us to observe a given target with the 
same or different filters simultaneously.  We observed \whya\ 
by using the filter pairs of (CntHa, NHa), (TiO717, Cnt748), and (Cnt820, 
Cnt820).  
For each target and with each filter pair, we took $N_{\rm exp}$ exposures 
for each of the Stokes \Qplus, \Qminus, \Uplus, and \Uminus\ components, 
with NDIT frames in each exposure. The polarization cycle of 
\Qplus, \Qminus, \Uplus, and \Uminus\ was repeated $N_{\rm pol}$ times. 
We carried out this procedure ($N_{\rm exp} \times N_{\rm pol}$) 
at three different dithering positions.  

The Strehl ratios of the observations of \whya\ were estimated from the 
$H$-band Strehl ratios from the auxiliary files of SPHERE observations 
(so-called GEN-SPARTA data) as described in Paper~I.  The Strehl ratios 
of the observations of the PSF reference HIP67834 were 
directly measured from the observed intensity maps. 
The Strehl ratios of the observations of \whya\ are 0.38 (CntHa/NHa), 
0.46 (TiO717/Cnt748), and 0.52 (Cnt820/Cnt820), thanks to good seeing 
(0.7--0.8\arcsec).  
For the PSF reference, the values are lower, 0.25 (CntHa/NHa), 
0.36 (TiO717/Cnt748), 
and 0.38 (Cnt820/Cnt820), because its flux at the wavelengths of 
the filters used in our observations is lower than that of \whya. 
The summary of our SPHERE-ZIMPOL observations is given in 
Table~\ref{obs_log_sphere}.

We reduced the SPHERE-ZIMPOL data using the pipeline 
version~0.15.0-2\footnote{Available at
  ftp://ftp.eso.org/pub/dfs/pipelines/sphere} in the 
same manner as described in Paper~I and obtained the maps of 
total (i.e., polarized + unpolarized) intensity ($I$), polarized intensity 
(\IP), degree of linear polarization (\PL), 
and position angle of the polarization vector. 
In Paper~I, we flux-calibrated the intensity maps using 
the visible spectrum of \whya\ taken at a variability phase close to the 
SPHERE-ZIMPOL observations. However, we could not find a spectrum 
of \whya\ covering the wavelengths of our SPHERE observations obtained near 
minimum light. Therefore, in the present work, we adopted the following 
approach. 
We first approximate the spectrum of the PSF reference star with a template 
spectrum with the corresponding effective temperature and the luminosity class 
available in the spectral library of Pickles (\cite{pickles98}). 
Based on the effective temperature of 4548~K estimated for HIP67834 
(McDonald et al. \cite{mcdonald12}), we adopted the spectrum of a K2III star 
in the spectral library of Pickles (\cite{pickles98}), whose effective 
temperature is the closest to HIP67834.  We reddended the spectrum with 
$A_V$ = 0.24, which was obtained using the parameterization of the interstellar 
extinction of Arenou et al. (\cite{arenou92}) 
and the wavelength-dependence of the interstellar 
extinction derived by Cardelli et al. (\cite{cardelli89}). 
Then the reddened template spectrum was scaled so that the flux computed with 
the $V$ band filter is equal to 7.64, which is the $V$ magnitude from Simbad. 
The resulting, flux-calibrated template spectrum allows us to compute the 
flux of our PSF reference star HIP67834 
with the SPHERE-ZIMPOL filters of our observations. 
Once the flux of HIP67834 with the SPHERE-ZIMPOL filters is known, 
the images of \whya\ can be flux-calibrated. 
The derived fluxes of \whya\ are listed in Table~\ref{whya_photometry}, 
together with the fluxes obtained at the first epoch in Paper~I.

\begin{figure*}[!hbt]
\begin{center}
\resizebox{\hsize}{!}{\rotatebox{0}{\includegraphics{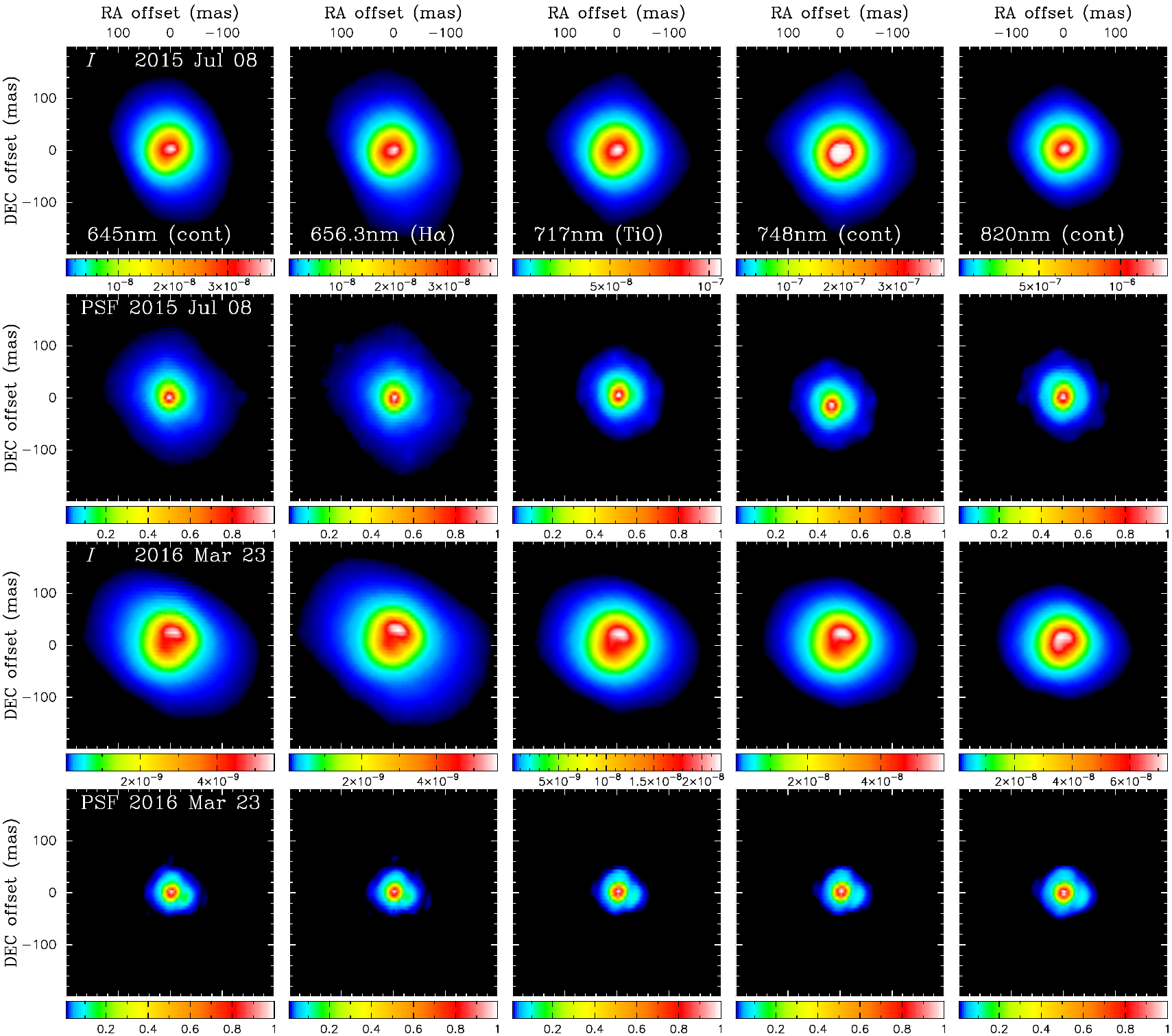}}}
\end{center}
\caption{
Intensity maps of \whya\ observed with SPHERE-ZIMPOL. 
The first and third rows show the data of \whya\ obtained on 2015 July 8 
(phase 0.92, Paper~I) and on 2016 March 23 (phase 0.54), respectively. 
The second and fourth rows show the intensity maps of the PSF reference star 
observed in 2015 and 2016, respectively. 
The observed images at 645~nm (CntHa, continuum), 656.3~nm (NHa, \Ha), 
717~nm (TiO717, TiO band), 748~nm (Cnt748, continuum), and 820~nm (Cnt820, 
continuum) are shown from the left to the right. 
North is up, east to the left in all panels.  
While the intensity peak is approximately at the 
center in 2015, the peak is clearly off-centered in 2016. 
The Strehl ratios for \whya\ are comparable between 2015 and 2016, 
although those for the PSF references differ significantly between 2015 
and 2016 (Sect.~\ref{subsubsect_res_sphere_dust}). 
The intensity maps of \whya, which are 
flux-calibrated as described in Sect.~\ref{subsect_obs_zimpol}, are 
shown in units of W~m$^{-2}$~\micron $^{-1}$~arcsec$^{-2}$.  
The color scale of the intensity maps is cut off at 1\% of the intensity peak.  
}
\label{whya_zimpol_I}
\end{figure*}

\subsection{Long-baseline spectro-interferometric observations with
  VLTI/AMBER}
\label{subsect_obs_amber}

The near-IR (1.3--2.4~\micron) VLTI instrument AMBER 
(Petrov et al. \cite{petrov07}) 
allows us to achieve a spatial resolution of 3~mas (at 2~\micron) 
with the current maximum baseline of 140~m by 
combining three Unit Telescopes (UTs) or 1.8~m Auxiliary Telescopes (ATs). 
The AMBER instrument operates with 
three spectral resolutions, 35, 1500, and 12000.  
The highest spectral resolution of 12000 is sufficient to resolve 
individual atomic and molecular lines.  

AMBER measures visibility, closure phase (CP), and differential phase (DP).  
The visibility, which is the amplitude of the Fourier transform of the object's 
intensity distribution on the sky, contains information about the size and 
shape of the object.  While the Fourier phase is destroyed by the atmospheric 
turbulence, the closure phase, which is the sum of the Fourier phases on three 
baselines around a triangle formed by three telescopes, conserves the 
phase information of the Fourier transform of the object's image. 
Deviations of CP from 0 or 180\degr\ indicate asymmetry of the object.  
The DP represents the photocenter shift in spectral features with respect to 
the continuum.  
The AMBER instrument also records the spectrum of the same wavelength 
region simultaneously with the interferometric fringes. 

Our VLTI/AMBER observations of \whya\ took place on 2014 February 11 (UTC) 
with the AT configuration B2-C1-D0, which covered projected baseline lengths 
from 11.3 to 33.9~m (Program ID: 092.D-0461(A), P.I.: K.~Ohnaka).  
Our AMBER observations covered the wavelength region between 2.28 and 
2.31~\micron\ near the CO first overtone 2--0 band head with the spectral 
resolution of 12000.  
A summary of our AMBER observations is given in Table~\ref{obs_log_amber}. 
The variability phase at the time of our AMBER observations is estimated 
to be 0.59 from the AAVSO light curve (but in a different variability cycle, 
see Sect.~\ref{subsect_res_amber} for discussion about the possible 
cycle-dependence of the angular size), 
which is close to the phase at the time of our second-epoch SPHERE-ZIMPOL 
observations. 
We observed \alfcenA\ (G2V, $K = -1.5$, uniform-disk 
diameter = $8.314 \pm 0.016$~mas, Kervella et al. \cite{kervella03}) 
as an interferometric and spectroscopic calibrator.

The AMBER data were reduced using the amdlib 
ver~3.0.7\footnote{Available at 
http://www.jmmc.fr/data\_processing\_amber.htm}, 
which is based on the P2VM algorithm (Tatulli et al. \cite{tatulli07}; 
Chelli et al. \cite{chelli09}).  
Details of the reduction are described in Ohnaka et al. (\cite{ohnaka09}, 
\cite{ohnaka11}, and \cite{ohnaka13}). 
The wavelength calibration and the spectroscopic calibration of the \whya\ 
data were carried out with the method described in 
Ohnaka et al. (\cite{ohnaka13}).

\begin{figure*}[!hbt]
\begin{center}
\resizebox{\hsize}{!}{\rotatebox{0}{\includegraphics{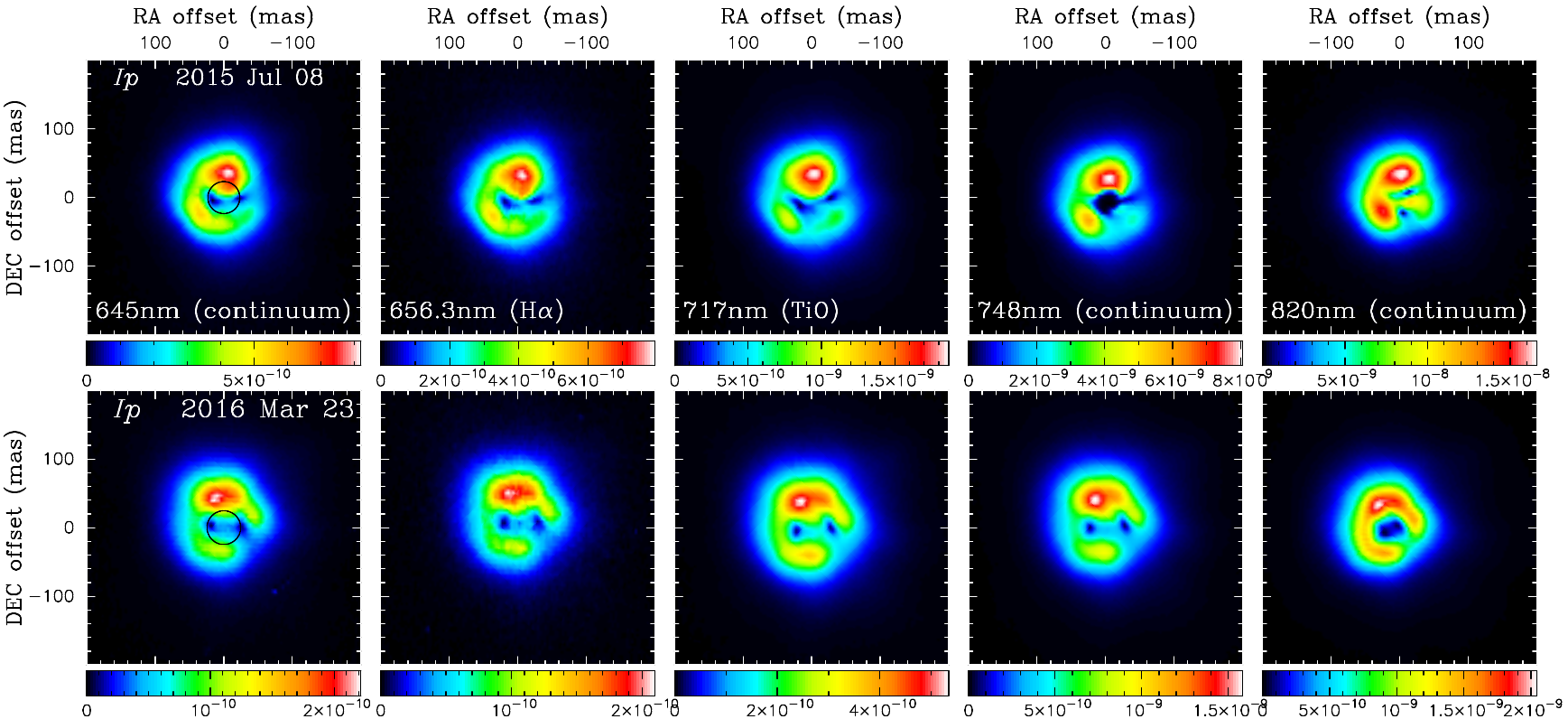}}}
\end{center}
\caption{
Polarized intensity maps of \whya\ observed with SPHERE-ZIMPOL. 
The top and bottom rows show the data of \whya\ obtained on 2015 July 8 
(phase 0.92, Paper~I) and on 2016 March 23 (phase 0.54), respectively. 
The observed images at 645~nm (CntHa, continuum), 656.3~nm (NHa, \Ha), 
717~nm (TiO717, TiO band), 748~nm (Cnt748, continuum), and 820~nm (Cnt820, 
continuum) are shown from the left to the right. 
North is up, east to the left in all panels.  
The 2016 data reveal the formation of a new, small dust clump in the west, 
while the dust clump detected in the SW in 2015 disappeared in 2016. 
The polarized intensity is 
shown in units of W~m$^{-2}$~\micron $^{-1}$~arcsec$^{-2}$.  
The circles in the leftmost panels represent the size 
of the star measured in the continuum near 2.3~\micron\ based on our 
VLTI/AMBER observations. 
}
\label{whya_zimpol_IP}
\end{figure*}

\section{Results}
\label{sect_res}

\subsection{SPHERE-ZIMPOL polarimetric images}
\label{subsect_res_sphere}

\subsubsection{Clumpy dust clouds}
\label{subsubsect_res_sphere_dust}

\begin{figure*}[!hbt]
\begin{center}
\resizebox{\hsize}{!}{\rotatebox{0}{\includegraphics{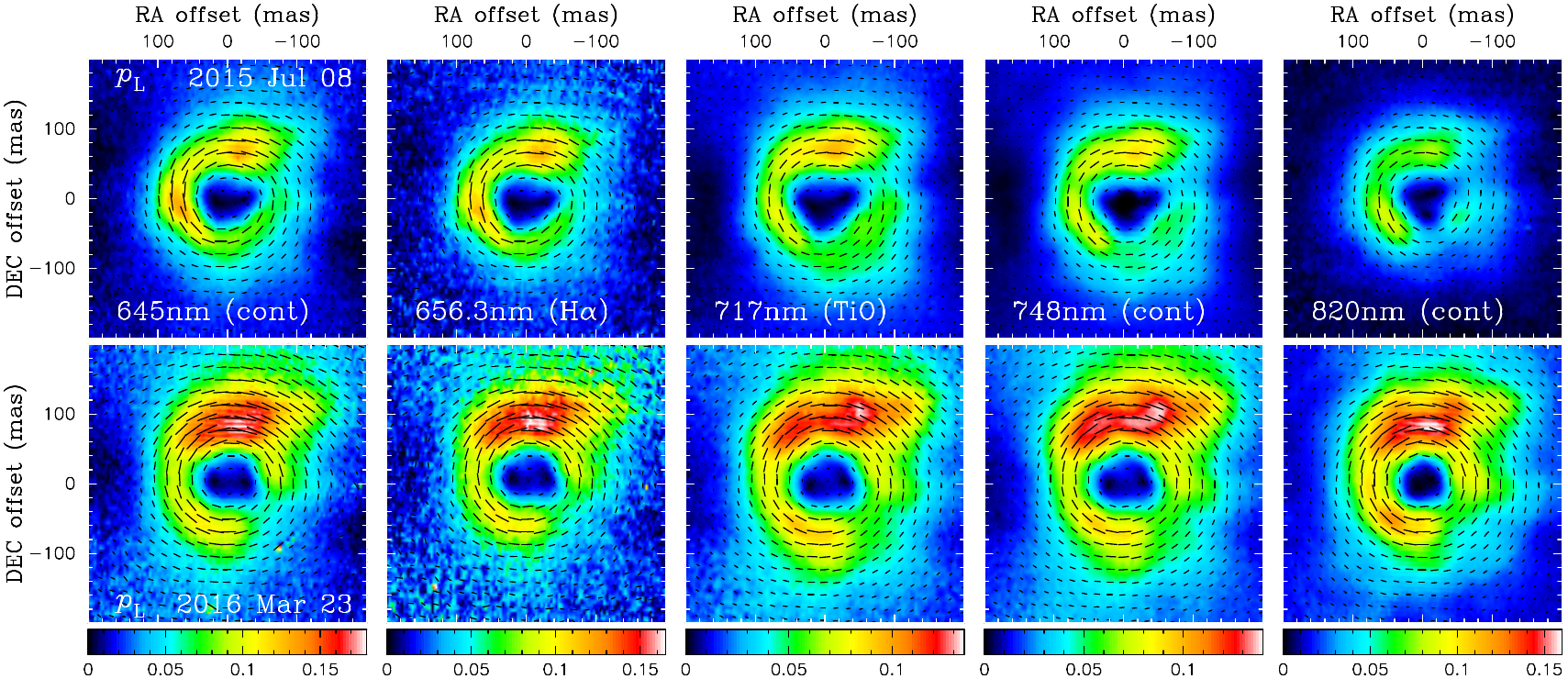}}}
\end{center}
\caption{
Degree of linear polarization maps of \whya\ observed with SPHERE-ZIMPOL. 
The top and bottom rows show the data of \whya\ obtained on 2015 July 8 
(phase 0.92, Paper~I) and on 2016 March 23 (phase 0.54), respectively. 
The images observed at 645~nm (CntHa, continuum), 656.3~nm (NHa, \Ha), 
717~nm (TiO717, TiO band), 748~nm (Cnt748, continuum), and 820~nm (Cnt820, 
continuum) are shown from the left to the right. 
At each wavelength, the same color scale is used for the 2015 and 2016 data, 
which makes it easier to recognize that the degree of polarization measured 
in 2016 is systematically higher than that measured in 2015. 
North is up, east to the left in all panels.  
}
\label{whya_zimpol_PL}
\end{figure*}

Figures~\ref{whya_zimpol_I}, \ref{whya_zimpol_IP}, and \ref{whya_zimpol_PL} 
show the maps of total intensity ($I$), 
polarized intensity (\IP), and degree of linear polarization (\PL) with the 
polarization vectors overlaid, respectively, obtained for \whya. 
In each figure, we included the data from Paper~I obtained at the first epoch 
(2015 July 8) corresponding to pre-maximum light (phase 0.92). 
In Fig.~\ref{whya_zimpol_I}, the intensity maps of the PSF references are 
shown. 
The 2-D Gaussian fit to the observed images of HIP67834 results in 
PSF FWHMs of $22\times24$~mas at 645 and 656.3~nm, $23\times26$~mas at 
717 and 748~nm, and $26\times27$~mas at 820~nm. 
We note that the Strehl ratios of the data of \whya\ are similar 
for the observations at the first and second epochs: 
(Strehl$_{2015}$, Strehl$_{2016}$) = (0.37, 0.38) for the CntHa/NHa pair, 
(0.51, 0.46) for the TiO717/Cnt748 pair, and (0.39, 0.52) for the 
Cnt820/Cnt820 pair. 
Therefore, our 
comparison of the images taken at two epochs is not severely affected by 
the difference in the AO performance. 

The intensity maps of \whya\ obtained at the second epoch 
(Fig.~\ref{whya_zimpol_I}, third row) are significantly more 
extended than the PSF references, indicating that the circumstellar envelope 
is spatially resolved.  
While the peak of the intensity maps 
obtained at the first epoch is approximately at the center 
of the images (Fig.~\ref{whya_zimpol_I}, top row), 
the peak of the images obtained at the second epoch is 
clearly off-centered. This off-centered peak can be a bright spot in the 
atmosphere of the central star or a bright dust clump. As described below, 
however, the polarized intensity maps, which reflects the spatial distribution 
of scattering material in optically thin cases, 
do not show a bright clump corresponding to the off-centered peak in the 
total intensity images. 
Therefore, it is more likely that it represents 
a bright spot in the atmosphere of the star. 
Similar inhomogeneities and their time variations within 48~days 
have recently been detected in another AGB star, R~Dor, 
by Khouri et al. (\cite{khouri16}). 
They interpret the inhomogeneities and their time variations as due to 
the fluctuations in the density and/or molecular abundance and/or 
excitation. 

The polarized intensity maps reveal more detailed structures close to 
the star with the unpolarized light from the bright central star effectively 
suppressed. 
The \IP\ maps shown in Fig.~\ref{whya_zimpol_IP} reveal three 
clumpy dust clouds: a large, elongated bright clump at 43~mas away 
in the north, a smaller clump at 34~mas in the south, and an even smaller 
clump at 50~mas nearly in the west. The three clumps appear to be connected 
on the \IP\ map observed at 820~nm. 
Given the star's angular diameter of $49.1 \pm 1.5$~mas measured with 
VLTI/AMBER (see Sect.~\ref{subsect_res_amber}), 
the location of the clumpy clouds 
correspond to 1.4--2.0~\RSTAR, revealing again dust formation very close to 
the star.  
Comparison with the \IP\ maps obtained at the first epoch in 2015 reveals 
that while the overall appearance 
of the clumpy dust clouds is similar (i.e., a large clump in the north and 
a smaller clump in the south), there are variations in the detailed 
structure.  On the one hand, the small clump detected in the west of the star 
at the second epoch is not seen in the data obtained at the first epoch, 
and therefore, it must have formed in the last 8.5 months.  On the other hand, 
the clump seen in the SW of the star in 2015 has nearly disappeared. 
The relative errors in the polarized intensity on the bright clumps are 
4--5\%, 7--10\%, 6--9\%, 5--8\%, and 4--6\% at 645, 656.3, 717, 748, and 
820~nm, respectively.

The FWHMs measured at 645, 656.3, 717, 748, and 820~nm at minimum light 
in 2016 are 68, 68, 72, 70, and 64~mas, respectively. 
These values are noticeably larger than the 
FWHMs measured in 2015 at pre-maximum: 
53~mas at 645~nm, 51~mas at 656.3~nm, 58~mas at 717~nm, and 46~mas at 820~nm 
(the central region of the 2015 image at 748~nm was affected by saturation). 
Therefore, 
our two-epoch observations suggest that the object (including the star 
itself, extended atmosphere, and dust clouds) appeared more extended at 
minimum light than at pre-maximum light. 
Moreover, the above FWHMs measured at minimum light are 
significantly larger than the values obtained by 
Ireland et al. (\cite{ireland04})---75~mas (717~nm), 50~mas (748~nm), and 
42~mas (820~nm)---except for 717~nm, although they observed near minimum 
light (phase 0.44). Given that the size measured in the visible 
is significantly affected by the dust clouds, the difference between the 
FWHMs measured in our present work and Ireland et al. (\cite{ireland04}) 
implies significant cycle-dependence of dust formation.

\begin{figure*}[!hbt]
\begin{center}
\resizebox{\hsize}{!}{\rotatebox{0}{\includegraphics{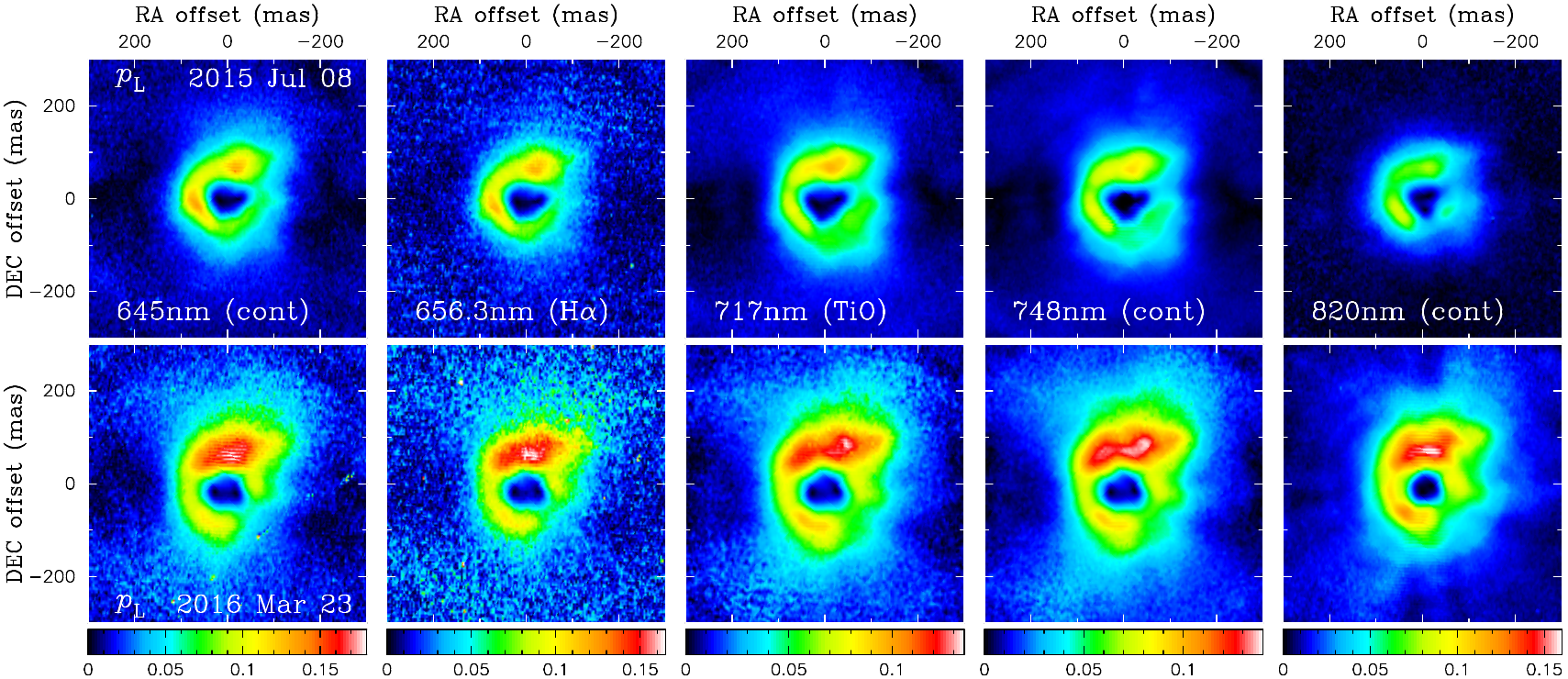}}}
\end{center}
\caption{
Degree of linear polarization maps of \whya.  
Same as Fig.~\ref{whya_zimpol_PL} but 
with a larger field of view to show the entire extent of the polarization 
signals, which is noticeably more extended in 2016 than in 2015. 
}
\label{whya_zimpol_PL_largeFOV}
\end{figure*}

The \PL\ maps shown in Fig.~\ref{whya_zimpol_PL} are characterized by 
an incomplete, asymmetric ring, which suggests an approximately shell-like 
distribution of dust. This is supported by the concentric pattern of 
the polarization vector maps.  The maximum degree of 
linear polarization measured in 2016 is 18\%, 16\%, 13\%, 14\%, and 16\% 
at 645, 656.3, 717, 748, and 820~nm, respectively. 
The absolute errors in the degree of linear polarization are 
0.8\% (i.e., 18\% $\pm$ 0.8\%), 1\%, 1\%, 0.7\%, and 0.6\% at 645, 656.3, 
717, 748, and 820~nm, respectively, in the region within $\sim$180~mas. 
In the region beyond this, the absolute errors are higher, 2.5\%, 5\%, 3\%, 
2\%, and 1.3\% at 645, 656.3, 717, 748, and 820~nm, respectively. 
The maximum degree of polarization measured in 2016 are systematically higher 
than those measured in 2015: 13\%, 12\%, 10\%, 9\%, and 8\% at 645, 656.3, 
717, 748, and 820~nm, respectively. 
Furthermore, as can be seen in Fig.~\ref{whya_zimpol_PL_largeFOV}, 
while the polarization was detected only up to $\sim$150~mas from 
the central star in 2015, it extends to $\sim$250~mas in 2016. 
This indicates that the distribution of 
the scattering material is more extended at phase 0.54 in 2016 than 
at phase 0.92 in 2015. 

This can be interpreted as the consequence of dust formation 
in a more extended region in 2016 than in 2015 or the result of the 
outward motions of the dust clouds detected in 2015. 
However, the angular displacement expected over the time interval of 
8.5 months between 
our two-epoch observations from the expansion velocity $\varv_{\rm exp}$~(\KMS) 
and the distance of 78~pc is $1.9 \times \varv_{\rm exp}$~mas. 
Since $\varv_{\rm exp}$ is expected to be 
smaller than the terminal velocity of 7.5~\KMS\ 
(Khouri et al. \cite{khouri15}), 
the angular displacement is smaller than 14~mas. This means that we cannot 
explain the detection of polarization in a more extended region in 2016 
by means of the expansion of the dust clouds detected in 2015. 
Therefore, it is more likely that the dust formation had taken place at larger 
radii at the time of our second-epoch observations 
compared to the first-epoch observations. 

\begin{figure*}
\begin{center}
\resizebox{\hsize}{!}{\rotatebox{0}{\includegraphics{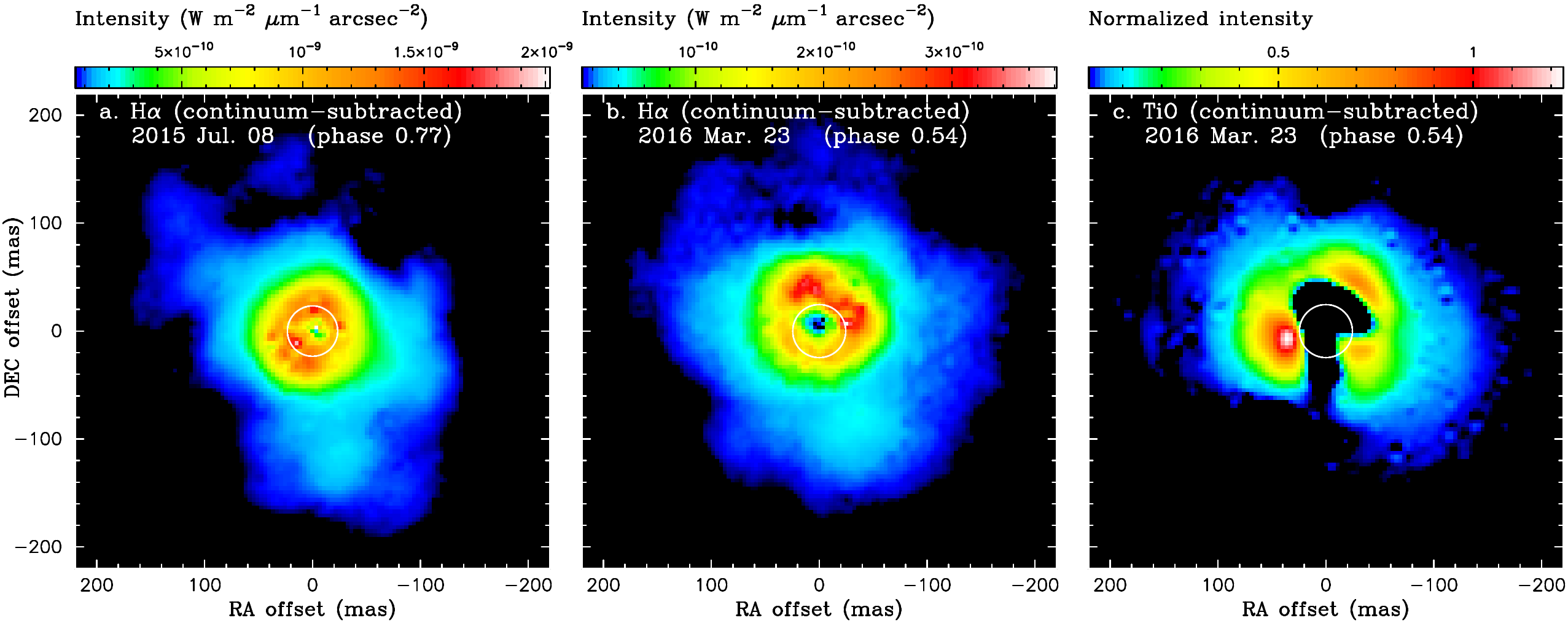}}}
\end{center}
\caption{
{\bf a:} Continuum-subtracted \Ha\ image of \whya\ derived from the 
SPHERE-ZIMPOL observations at pre-maximum light (phase 0.92) on 
2015 July 08 (Paper~I) . 
{\bf b:} Continuum-subtracted \Ha\ image of \whya\ observed at minimum light 
(phase 0.54) on 2016 March 23. 
{\bf c:} Continuum-subtracted TiO (717~nm) image of \whya\ observed 
at minimum light (phase 0.54) on 2016 March 23. 
North is up, east to the left.  
}
\label{whya_zimpol_Ha_TiO}
\end{figure*}

\subsubsection{Extended \Ha\ and TiO emission}
\label{subsubsect_res_sphere_Ha}

The simultaneous observations of the filter pairs (CntHa, NHa) and 
(TiO717, Cnt748) allow us to subtract the continuum image to study 
the extended \Ha\ and TiO emission in detail, because the AO performance 
was the same for each pair. 
We subtracted the continuum image taken with the CntHa filter from the NHa 
image, in which both images are flux-calibrated as described in 
Sect.~\ref{subsect_obs_zimpol}. 
Comparison of Figs.~\ref{whya_zimpol_Ha_TiO}a and \ref{whya_zimpol_Ha_TiO}b 
reveals that the images taken at two epochs show strong, nearly circular \Ha\ 
emission within $\sim$50~mas ($\sim$2~\RSTAR).  We detected fainter, 
more irregular-shaped emission extending to $\sim$200~mas ($\sim$8~\RSTAR) 
and $\sim$150~mas ($\sim$6~\RSTAR) at pre-maximum light in 2015 and 
at minimum light in 2016, respectively. 
The image taken in 2016 shows more emission in the E-W direction 
compared to the data taken in 2015.  

While the overall spatial extent of the \Ha\ emission is comparable 
at both epochs, 
there are time variations in the clumpy structures. For example, the 
clumpy emission detected at $\sim$125~mas ($\sim$5.4~\RSTAR) in the SSW 
in 2015 appears at an angular distance of $\sim$115~mas ($\sim$4.7~\RSTAR) 
at approximately the same position angle in 2016. 
The \Ha\ emission is considered to be associated with shocks 
induced by stellar pulsation (e.g., Fox \& Wood \cite{fox84}). 
In Paper~I, we interpreted the extended \Ha\ emission as originating from 
the shocks propagating to as far as $\sim$5~\RSTAR. 
However, as the shocks are significantly weakened at such large radii, 
it is more likely that the \Ha\ emission originates from the 
shocks much closer to the star and is scattered at large radii probably by 
dust.  Possibly the strong, nearly circular emission 
within $\sim$50~mas ($\sim$2~\RSTAR) may represent the direct emission 
from the shocks.  
In this scenario, the \Ha\ emission level is expected to change 
with pulsation phase, which is indeed observed, 
while the dependence on pulsation phase should be smeared out and be weak 
if the emission originates from the shocks far away from the star.

The \Ha\ emission seems to be more extended at the position angles of 
the gaps between the dust clouds seen in the \IP\ maps 
(Fig.~\ref{whya_zimpol_IP}). 
The \Ha\ emission in 2016 shows more extension to the east, SSW, and west, 
which approximately---though not exactly---corresponds to the gaps 
between the dust clouds. 
Similar anti-correlation is seen in the 2015 data as well. The extended \Ha\ 
emission toward the south, SWW, and NE seems to correspond to the cloud gaps, 
which is particularly clear in the \IP\ map at 717~nm 
(Fig.~\ref{whya_zimpol_IP}, upper row, third column). 
If the extended \Ha\ emission beyond $\sim$50~mas represents the \Ha\ 
photons that originate in the inner shocks and are scattered by dust grains 
at large radii, we can expect that the inner dust clouds can partially 
block the \Ha\ photons, leading to weaker emission at large radii in 
the directions of the dust clouds. 
Therefore, the anti-correlation between the inner dust clouds 
and the \Ha\ emission at large radii, albeit weak, is also consistent with the 
aforementioned origin of the extended \Ha\ emission.

To study the extended TiO emission in the 2016 data, 
we first subtracted the flux-calibrated image obtained with the Cnt748 filter 
from the TiO717 image.  However, the resulting 
image shows negative values in the entire region for the following reason. 
The TiO band is expected to appear in faint emission (positive pixel 
values) in the extended atmosphere (i.e., outside the limb of the star), while 
strong TiO absorption (negative pixel values) is expected over the stellar disk. 
When the strong TiO absorption over the stellar disk is convolved with 
the PSF, it dominates the faint extended emission, resulting in negative 
values in the entire image. Therefore, we subtracted the TiO717 and Cnt748 
images normalized with the intensity at the center of the stellar disk, 
which was assumed to be at the centroid of the nearly zero-polarization 
region over the central star (see Fig.~\ref{whya_zimpol_PL}).  
This method allows us to 
recognize the extended emission in positive pixel values and ascertain its 
spatial extent. 
On the other hand, because the absolute flux scale is lost, we cannot 
properly interpret the inhomogeneous absorption over the stellar disk, 
for example, possible correspondence between the inhomogeneous TiO absorption 
seen over the stellar disk and the off-centered bright spot detected in the 
intensity maps. 
Figure~\ref{whya_zimpol_Ha_TiO}c shows the continuum-subtracted TiO717 
image, which reveals the TiO emission extending to $\sim$150~mas 
($\sim$6~\RSTAR), overlapping with the extended \Ha\ emission shown 
in Fig.~\ref{whya_zimpol_Ha_TiO}b.

\begin{figure*}
\begin{center}
\resizebox{\hsize}{!}{\rotatebox{-90}{\includegraphics{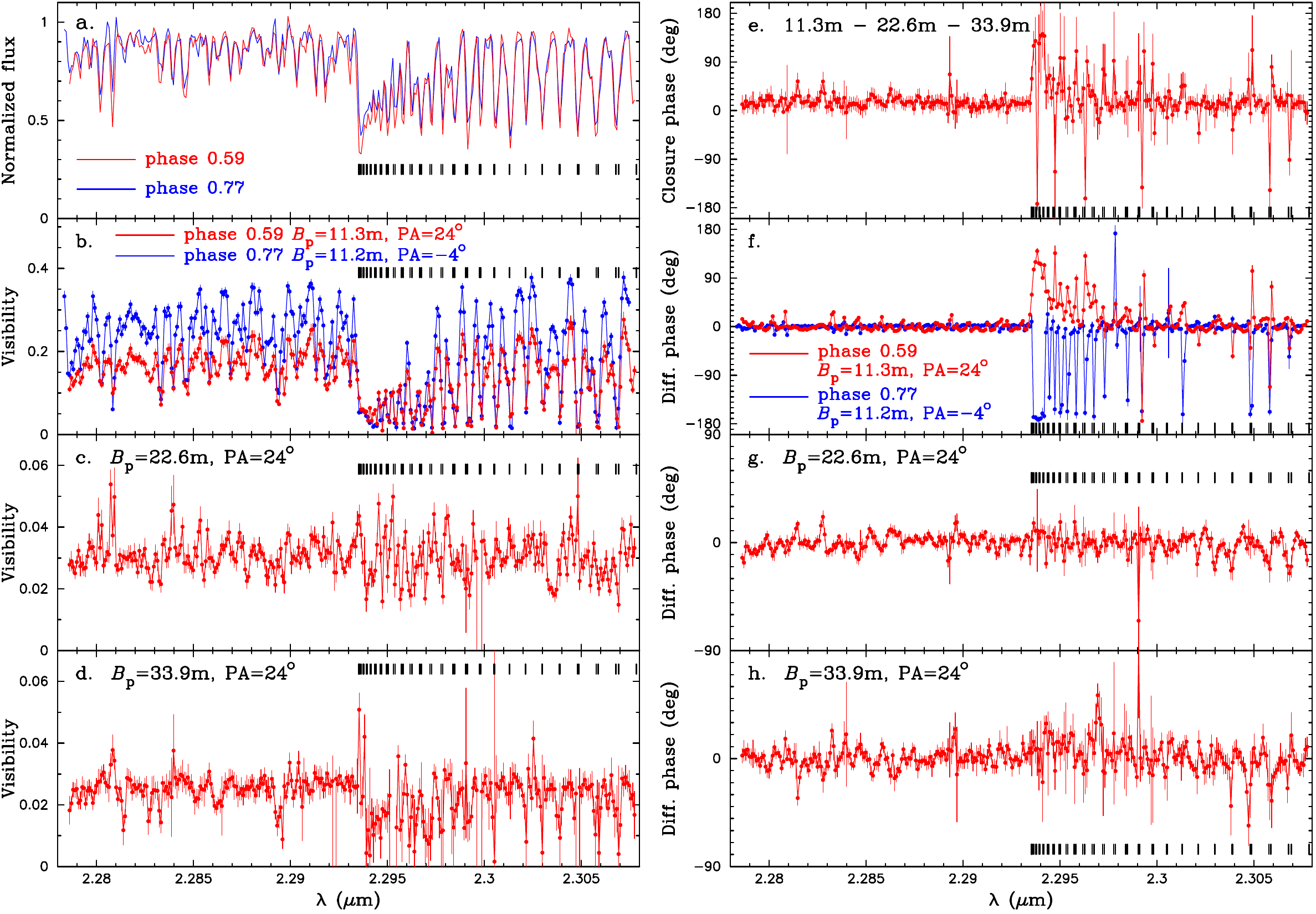}}}
\end{center}
\caption{
VLTI/AMBER observations of \whya\ with a spectral resolution of 12000 
near minimum light (phase 0.59) on 2014 February 11 
(data set \#3).  
{\bf a:} Observed spectrum. 
{\bf b}--{\bf d:} Visibilities. 
{\bf e:} Closure phase.  
{\bf f}--{\bf h:} Differential phases. 
The positions of the CO lines are indicated by the ticks. 
In the panels {\bf a}, {\bf b}, and {\bf f}, the data from Paper~I, 
taken at phase 0.77, are also shown. 
}
\label{whya_res_amber}
\end{figure*}

\subsection{AMBER observations of the central star and molecular outer
  atmosphere}
\label{subsect_res_amber}

Figure~\ref{whya_res_amber} shows the visibilities, DPs, and CP measured 
with VLTI/AMBER. The simultaneously observed spectrum is plotted in 
Fig.~\ref{whya_res_amber}a to 
show the correspondence between the signatures in the spectrum and in the 
interferometric observables.  The spectrum shows the salient CO first overtone 
lines longward of the band head at 2.293~\micron. In addition, a number of 
weak lines due to \HOH\ and CN are present shortward of the band head. 
Comparison of the spectra taken at two epochs 
suggests that the CO band head and the individual 
lines near the band head ($\la$2.302~\micron) are slightly deeper at the 
second epoch (minimum light) than at the first epoch (pre-maximum light). 
The weak \HOH\ and CN lines shortward of the CO band head do not indicate 
significant time variations (the lines at 2.28 and 2.281~\micron\ are 
affected by the residual of the removal of the telluric lines). 

The observed visibilities show clear 
signatures not only in the CO lines but also in the weak \HOH\ and CN lines. 
The projected baseline of 11.3~m of the present data is 
very close to one of the baselines of the data reported in Paper~I (11.2~m), 
obtained at phase 0.77. 
Therefore, the visibility measured at that baseline from Paper~I 
is overplotted in Fig.~\ref{whya_res_amber}c to show time variations. 
Obviously, the visibility shows more pronounced time variations than the 
spectra: the visibility obtained near minimum light (phase 0.59) is 
noticeably lower than that obtained at phase 0.77, 
suggesting that the object appears larger at minimum light than at a later 
phase closer to maximum light. 
This is qualitatively consistent with the time variations 
in the angular size of \whya\ measured from 1.1 to 3.8~\micron\ by 
Woodruff et al. (\cite{woodruff09}). 
We note, however, that the position angle of the projected baseline differed 
by 28\degr\ at two epochs.  Therefore, it is possible that deviation from 
circular symmetry is at least partially responsible for the observed 
difference in the visibility. 

\begin{figure}[!hbt]
\begin{center}
\resizebox{\hsize}{!}{\rotatebox{0}{\includegraphics{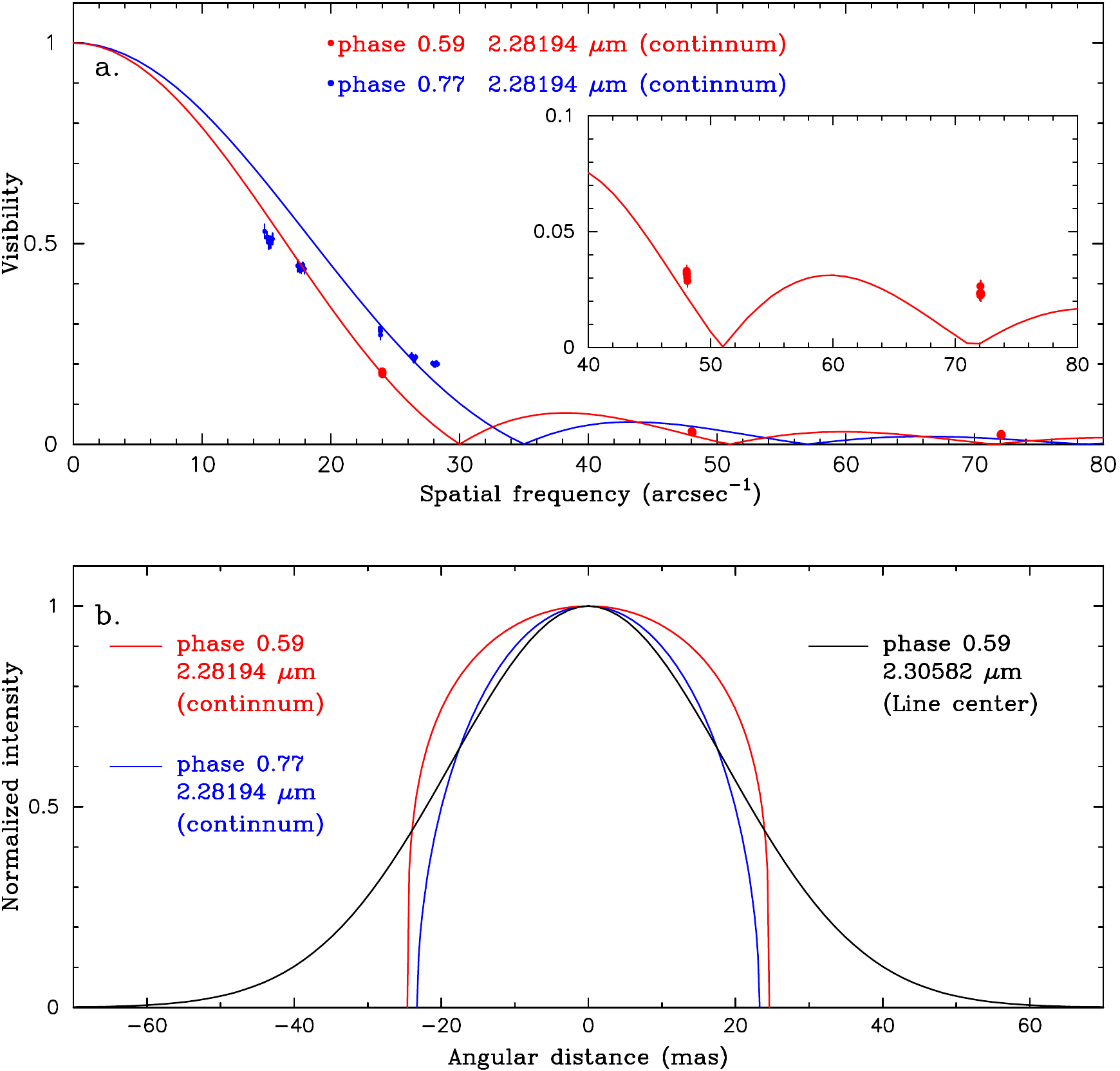}}}
\end{center}
\caption{
Limb-darkened disk fit to the AMBER data of \whya.  
{\bf a:} Visibilities observed at a continuum wavelength on 2014 February 11 
(phase 0.59) are plotted with the red dots, together with the power-law-type 
limb-darkened disk fit (red solid line).  The inset shows an enlarged view 
of the second, third, and fourth visibility lobes. 
The visibilities observed at the 
corresponding wavelength on 2014 April 22 (phase 0.77) 
are shown by the blue dots, together with the limb-darkened disk fit 
(blue solid line). 
{\bf b:} Limb-darkened disk intensity profiles derived for phase 0.59 
(red solid line) and 0.77 (blue solid line) at the continuum wavelengths 
shown in the panel {\bf a}. The Gaussian intensity profile in one of the 
CO lines obtained at phase 0.59 is plotted by the black solid line. 
}
\label{whya_amber_lddfit}
\end{figure}

To estimate the angular size of the star, we fitted the observed visibilities 
with a power-law-type limb-darkened disk 
(Hestroffer et al. \cite{hestroffer97}). 
Then we computed the average of the limb-darkened disk diameter 
and limb-darkening parameter derived at the continuum wavelengths 
shortward of the band head, avoiding 
the moderately strong lines.  
Figure~\ref{whya_amber_lddfit}a shows the fit at one of the continuum 
wavelengths, together with the fit at the same wavelength for the 2015 data. 
The averaged limb-darkened disk diameter 
and limb-darkening parameter in the continuum are $49.1 \pm 1.5$~mas and 
$1.16 \pm 0.49$, respectively. 
For the data taken at phase 0.77, 
we obtained the average limb-darkened disk diameter of $46.6 \pm 0.1$~mas 
and the limb-darkening parameter of $2.1\pm0.2$ (Paper~I). 
Therefore, as can be seen in Fig.~\ref{whya_amber_lddfit}b, 
the star appears larger and shows weaker limb-darkening near minimum light 
than at pre-maximum light. 
However, Fig.~\ref{whya_amber_lddfit}a reveals that the data show clear and 
significant discrepancy from the limb-darkened disk fit. The median of the 
reduced $\chi^2$ at the continuum wavelengths is 29.9, which also suggest 
that the star shows significant deviation from a limb-darkened disk. 

We note that the AMBER observations in Paper~I (2014 April 22) 
and the present paper (2014 February 11) 
were carried out in the same variability cycle of \whya.  
Therefore, the cycle-to-cycle variation is not responsible for the time 
variations in the AMBER data. 
On the other hand, 
as mentioned in Sect.~\ref{subsect_obs_amber}, our AMBER observations of 
\whya\ took place at a phase close to that of the SPHERE/ZIMPOL observations 
but in a different variability cycle.  We estimate, however, that the 
cycle-dependence of the angular size near 2.3~\micron\ is small based on 
the aperture-masking observations of \whya\ over eight variability cycles 
($\sim$8.3~years) by Woodruff et al. (\cite{woodruff08}). 
The $K$-band (2.26~\micron) uniform-disk diameters that they derived near 
minimum light (instead of a limb-darkened disk as we used) 
range from 41.5 to 45.9~mas. This suggests a cycle-to-cycle variation in 
the uniform-disk diameter of $\pm$2.2~mas.  If we adopt this value for 
the cycle-to-cycle variation in the limb-darkened disk, 
the estimate of the distance of the clumpy dust clouds in terms of the stellar 
radius is affected slightly, by $\pm$5\% at most. Therefore, it does not 
significantly change the interpretation of the results.

For the CO lines, we could not achieve any reasonable fit to the observed data 
with a limb-darkened disk. 
The second and third baselines (projected baseline 
lengths of 22.6 and 33.9~m) correspond to the second and third visibility lobe 
in the continuum.  Since the star appears more extended in the CO lines, 
these baselines may correspond to even higher visibility lobes in the 
CO lines. 
The visibilities at such high lobes are affected by inhomogeneities, 
whose presence can also be inferred from the non-zero DPs and 
non-zero/non-180\degr\ CP (Figs.~\ref{whya_res_amber}e--h). 
Probably these inhomogeneities make the fit with a simple, symmetric 
limb-darkened disk impossible. 
Nevertheless, to obtain a rough estimate of the spatial extent of the 
atmosphere in the CO lines, we derived the Gaussian FWHM only from the 
visibilities at the shortest baseline.  The FWHMs in the CO lines range 
from 40 to 46~mas, and the Gaussian intensity profile in one of the CO lines 
(at the line center) is plotted in Fig.~\ref{whya_amber_lddfit}b.  
The figure suggests that the atmosphere appears more than twice as extended 
in the CO lines as in the continuum.

\section{Determination of luminosity and effective temperature at 
pre-maximum and minimum light}
\label{sect_stellar_param}

In Paper~I, we adopted an effective temperature of 2500~K based on the 
modeling of the spectral energy distribution (SED) of \whya\ 
carried out by Khouri et al. (\cite{khouri15}) and derived a luminosity 
of 5130~K by combining the effective temperature with the angular diameter 
measured with VLTI/AMBER and the distance of 78~pc. 
However, now that we have SPHERE-ZIMPOL data taken at two distinct phases, 
it is necessary to use the effective temperature and luminosity appropriate 
for each epoch.  

To derive the bolometric flux at the second epoch (minimum light), 
we collected (spectro)photometric data as follows. 
We used the flux derived with five SPHERE-ZIMPOL filters 
(Table~\ref{whya_photometry}) in the visible. 
For the near-IR domain, we scaled the spectrum 
observed at phase 0.58 by Woodruff et al. (\cite{woodruff09}, see their 
Fig.~4) to an $H$-magnitude of $-2.4$, which is a typical value near 
minimum light based on the long-term photometry by Whitelock et al. 
(\cite{whitelock00}). In the wavelength region longward of 2.4~\micron, 
we adopted the spectrum obtained with the Short Wavelength Spectrometer 
(SWS) onboard the Infrared Space Telescope (ISO). We downloaded the 
spectrum observed on 1997 January 7 near minimum light (phase 0.56) 
from the ISO data 
archive\footnote{http://www.cosmos.esa.int/web/iso/access-the-archive}. 
The collected (spectro)photometric data were dereddened by $A_V = 0.245$ 
derived using the parameterization of the interstellar extinction of 
Arenou et al. (\cite{arenou92}). The bolometric flux was derived to be 
$1.69\times10^{-8}$~W~m$^{-2}$. This translates into a bolometric luminosity 
of $3180^{+550}_{-440}$~\LSOL\ when combined with the distance of 
$78^{+6.5}_{-5.6}$~pc. The derived bolometric flux and the angular diameter 
of 49.1~mas measured in the continuum near 2.3~\micron\ with VLTI/AMBER 
lead to an effective temperature of 2140~K. 

We also derived the luminosity and the effective temperature at 
the first-epoch (pre-maximum, phase 0.92) to check whether the values 
adopted in Paper~I can be justified or not. 
Instead of the flux derived with the SPHERE-ZIMPOL filter, we used the 
high-resolution visible spectrum presented by 
\mbox{Uttenthaler} et al. (\cite{uttenthaler11}) with our absolute flux 
calibration 
described in Paper~I. For the near-IR domain, we scaled the spectrum 
obtained at phase 0.79 by Woodruff et al. (\cite{woodruff09}, see their 
Fig.~5) with an $H$-magnitude of $-2.6$, which is a value representative 
of pre-maximum light based on the aforementioned study of 
Whitelock et al. (\cite{whitelock00}). 
For the wavelength region longward of 2.4~\micron, 
no ISO/SWS spectrum near pre-maximum is available. Therefore, we 
scaled the ISO/SWS spectrum taken on 1996 February 14 
at phase 0.67 to match the above flux-calibrated spectrum of 
Woodruff et al. (\cite{woodruff09}) at 2.4~\micron. 
The bolometric flux was found to be 
$2.40 \times 10^{-8}$~W~m$^{-2}$, which translates into a luminosity of 
$4520^{+780}_{-630}$~\LSOL\ with the distance of $78^{+6.5}_{-5.6}$~pc. 
The combination of this bolometric flux and the angular diameter of 
46.6~mas measured in the continuum in Paper~I results in an effective 
temperature of 2400~K. These luminosity and effective temperature 
agree reasonably well with the 5130~\LSOL\ and 2500~K adopted in Paper~I.

\begin{figure*}
\sidecaption
\rotatebox{0}{\includegraphics[width=12cm]{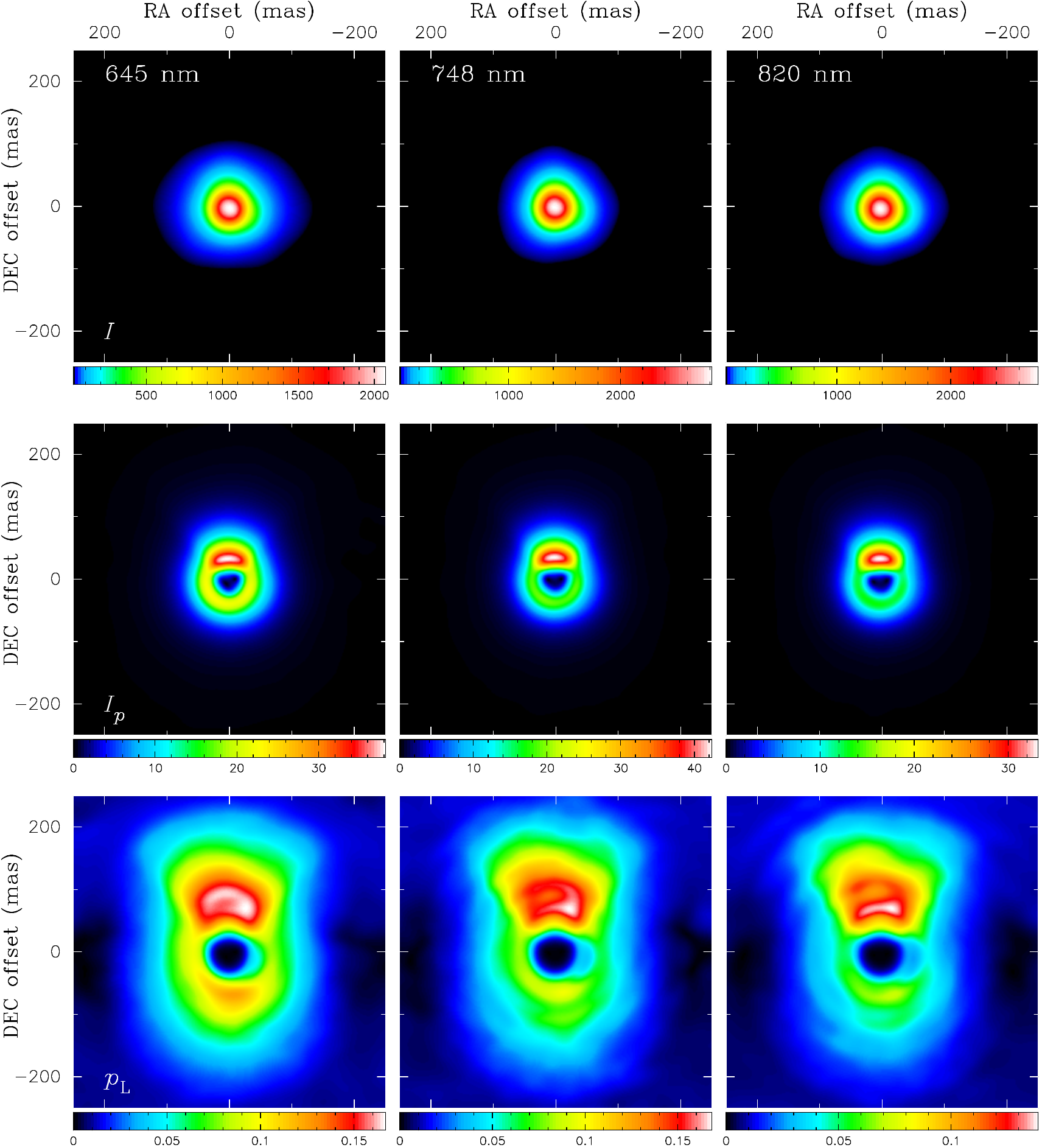}}
\caption{
Dust clump model for \whya.  
The top, middle, and bottom rows show the maps of the intensity, polarized 
intensity, and degree of linear polarization predicted by the model 
with the convolution with the observed PSF. 
The model images predicted at 645, 748, and 820~nm are shown in the 
first, second, and third columns. 
North is up, east to the left.  
}
\label{whya_model_images}
\end{figure*}

\begin{figure*}
\begin{center}
\resizebox{\hsize}{!}{\rotatebox{-90}{\includegraphics{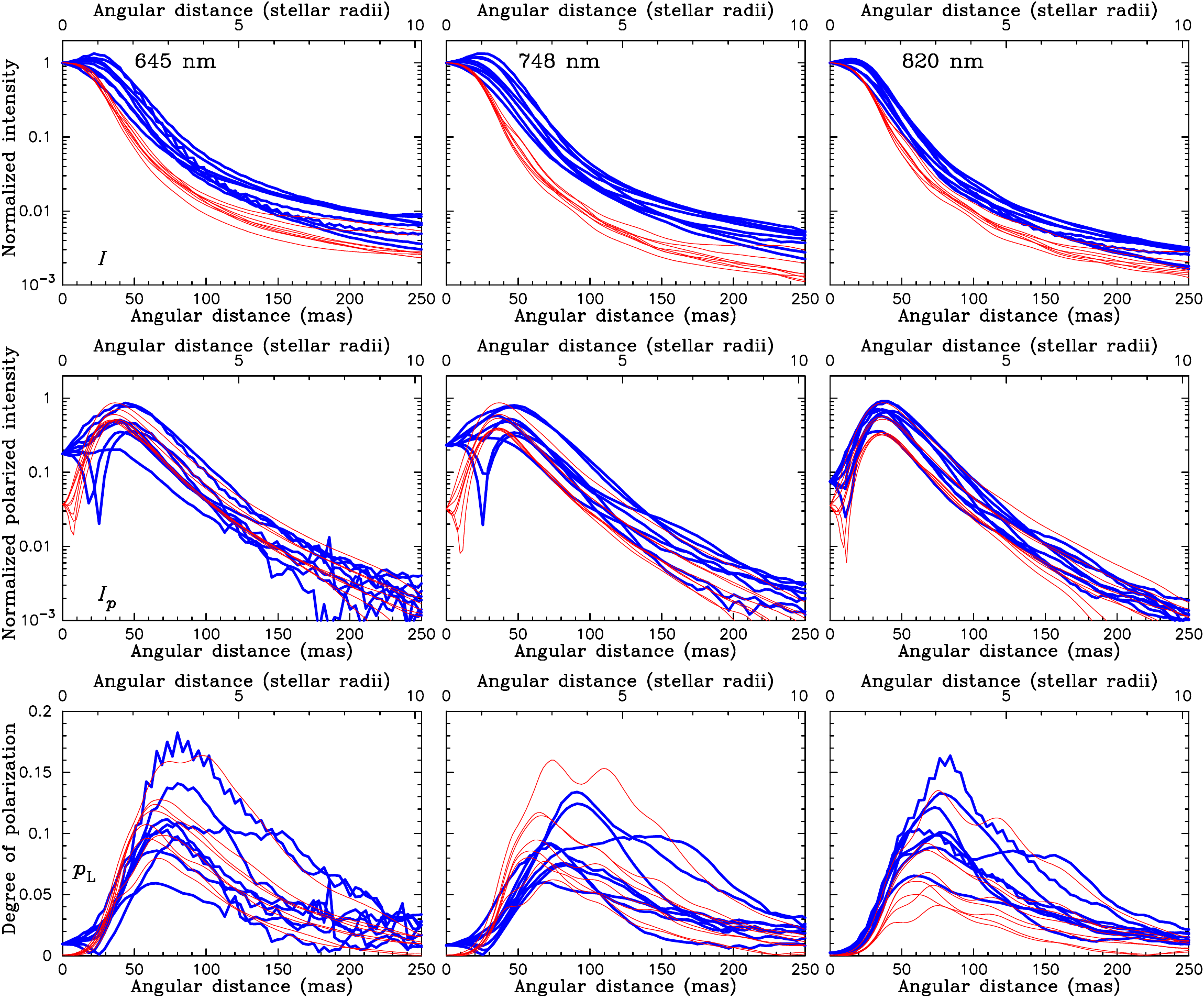}}}
\end{center}
\caption{
Comparison of the dust clump model and the observed SPHERE-ZIMPOL data. 
The top, middle, and bottom rows show the 1-D cuts at eight different position 
angles (0\degr, 45\degr, 90\degr, ..., 315\degr) of the intensity, normalized 
polarized intensity, and degree of polarization maps, respectively. 
The first, second, and third columns show the comparison at 645, 748, and 
820~nm, respectively. 
In each panel, the observed data are plotted with the thick blue lines, 
while the model is plotted with the thin red lines. 
}
\label{whya_model_1D}
\end{figure*}

\section{Monte-Carlo radiative transfer modeling}
\label{sect_modeling}

To derive the physical properties of the dust environment close to the star, 
we used the multi-dimensional dust radiative transfer code mcsim\_mpi 
(Ohnaka et al. \cite{ohnaka06}), adopting the shell with a cone-shaped 
density enhancement used in Paper~I. 
The radiation from the central star was approximated with the blackbody 
with 2140~K as derived in the previous section. 
The dust shell is defined by its inner and outer boundary radii.
The inner boundary radius is defined so that the dust temperature reaches 
a sublimation temperature of 1500~K. 
The dust density was assumed to be $\propto r^{-p}$, and 
$p$ was fixed to 3 in Paper~I, which approximately represents the density 
distribution in the wind acceleration 
region (see discussion in Sect.~4 of Paper~I). 
However, as Fig.~\ref{whya_zimpol_PL_largeFOV} shows, the dust distribution 
is more extended in 2016 than observed in 2015, which suggests that the 
optical depth is higher and/or the density gradient is shallower in 2016 
than derived from the 2015 data. Therefore, in the present work, we treated 
$p$ as a free parameter. 

The clumpy dust clouds seen in the SPHERE-ZIMPOL data are approximated with 
a cone-shaped density enhancement in a certain direction 
(see Fig.~8a of Paper~I). 
We adopted the half-opening angle of the density enhancement of 45\degr\ 
and a density ratio of 4 as derived in Paper~I, 
because the overall structure of the inner dust clouds seen in 2016 is similar 
to what was seen in 2015. For the dust opacities, we computed the absorption 
and scattering coefficients as well as the scattering matrix elements for 
spherical grains using the code of Bohren \& Huffman (\cite{bohren83}) 
from the complex refractive indices of \alumina\ measured 
by Koike et al. (\cite{koike95}, their ``Alumina'' sample), 
\enstatite, and \forsterite\ measured by J\"ager et al. (\cite{jaeger03}). 
The free parameters of our modeling are the optical depth of the dust shell 
in the radial direction, the outer radius of the dust shell, the exponent 
of the density distribution $p$, and the grain size.  
We compared the model images convolved with the observed PSFs 
with the maps of the intensity and the degree of polarization 
observed at 645, 748, and 820~nm, as well as the polarized intensity 
normalized with its peak as a byproduct. 
We avoided 656.3~nm and 717~nm, because the intensity 
maps at these wavelengths are affected by the extended \Ha\ and TiO emission, 
respectively, as described in Sect.~\ref{subsubsect_res_sphere_Ha}. 

Figure~\ref{whya_model_images} shows the best-fit model, 
which is characterized by a 550~nm optical 
depth of 0.6, an inner radius of 1.3~\RSTAR, an outer radius of 
10~\RSTAR, a density distribution exponent of 3, and 
\corundum\ grains with a size of 0.1~\micron. 
The dust mass of this model is $3.7 \times 10^{-9}$~\MSOL\ with a bulk density 
of 4~g~cm$^{-3}$ adopted for \corundum. 
The uncertainty in the 550~nm optical depth, outer radius, and density 
distribution exponent is $\pm 0.2$, $\pm 2$, 
and $\pm 0.5$, respectively. 

Figure~\ref{whya_model_1D} shows the 1-D profiles of the observed and model 
images at eight different position angles (0\degr, 45\degr, 90\degr, ..., 
315\degr). 
To avoid the bright, off-centered spot in the observed intensity maps, 
the center of the images was assumed to be the centroid of the 
nearly zero-polarization region over the central star. 
Figure~\ref{whya_model_1D} (top row) shows that 
the intensity maps predicted by the model are narrower than the observed 
data, particularly at 645 and 748~nm.  The bright, off-centered spot 
seen in the observed intensity maps makes the 1-D cuts broader, which could 
partially explain the discrepancy between the predicted intensity maps and 
the observed data. However, the discrepancy at 645 and 748~nm seems to be 
difficult to explain by this effect alone. 
The discrepancy may primarily be attributed to the extended TiO emission, 
as presented in Fig.~\ref{whya_zimpol_Ha_TiO}c.
The wavelengths sampled with the CntHa, Cnt748, and Cnt820 filters are 
not entirely free of the effects of the TiO bands, because the TiO bands 
are dominant in the visible.  Therefore, the extended TiO emission may be 
present at 645 and 748~nm, which can make the intensity 1-D cuts broader. 
On the other hand, 
the observed normalized polarized intensity and the degree of polarization 
are reproduced reasonably well.  
The shape of the polarized intensity profiles and the absolute values of 
the degree of polarization agree with the observed data. 
Moreover, given the simplicity of the model, the observed position angle 
dependence of the polarized intensity and the degree of polarization is 
also reproduced well.

We found out that the models with grain sizes larger than 0.2~\micron\ 
predict the degree of polarization at 645~nm to be too low compared to 
the observations. 
The models with grains smaller than 0.05~\micron\ also predict the degree of 
polarization to be too low. 
In these models, the amount of scattered light is much smaller, because 
the albedos are much lower for grains smaller than 0.05~\micron. 
The decrease in scattered light means 
a larger fraction of the unpolarized direct star light. Therefore, after 
convolving with the PSFs, the degree of polarization becomes lower. 
As in the case of Paper~I, the models with \forsterite\ and 
\enstatite\ also reproduce the observed data, and therefore, 
we cannot constrain the grain species from our SPHERE-ZIMPOL data alone. 

In Paper~I, we also used the fraction of scattered light derived from 
the near-IR polarimetric interferometric observations of Norris et al. 
(\cite{norris12}) as observational constraints. 
The fraction of scattered light computed from the model 
shown in Figs.~\ref{whya_model_images} and \ref{whya_model_1D}
is 0.06, 0.03, and 0.004 at 1.04, 1.24, and 2.06~\micron, respectively, 
which is much lower than the observed values of $0.176\pm0.002$ 
(1.04~\micron), $0.110\pm0.003$ (1.24~\micron), and $0.022\pm0.004$ 
(2.06~\micron). However, the observations of Norris et al. (\cite{norris12}) 
took place at phase 0.2, very different from the 0.54 of our second-epoch 
SPHERE-ZIMPOL observations.  Given the noticeable time variations seen in 
the observed maps of the polarized intensity and degree of polarization 
between phase 0.92 and 0.54 (Figs.~\ref{whya_zimpol_IP}, \ref{whya_zimpol_PL}, 
and \ref{whya_zimpol_PL_largeFOV}), 
it would not be surprising that the fraction of scattered light changes 
significantly between phase 0.2 and 0.54. Therefore, we did not attempt to 
reproduce the results of Norris et al. (\cite{norris12}) in the present 
modeling. 

Our modeling of the 2015 data presented in Paper~I suggested the 
predominance of grains as large as 0.5~\micron. To check whether or not the 
presence of such large grains can be excluded for the 2016 data, 
we computed models with 0.1~\micron\ and 0.5~\micron\ \alumina\ grains.  
For simplicity, 
we set the inner and outer radii of the distribution of 
two grain species as well as 
the exponent of the density distribution to be the same as those 
of the best-fit model with 0.1~\micron\ \alumina\ grains alone. 
It turned out that the models with \tauV\ = 0.8 (0.1~\micron\ grains) 
and \tauV\ $\le$ 0.1 (0.5~\micron\ grains) can fairly reproduce the 
SPHERE-ZIMPOL data.  
These optical depths correspond to dust mass of 
$4.9\times10^{-9}$~\MSOL\ and $\le 5.5\times10^{-10}$~\MSOL\ for 
0.1~\micron\ and 0.5~\micron\ grains, respectively. 
The 550~nm optical depth of the 0.5~\micron\ 
grains in the best-fit model for the 2015 data is 0.1. 
Therefore, we cannot exclude the presence of 0.5~\micron\ grains with 
an amount comparable to what was found in 2015.

\begin{table}
\begin{center}
\caption {
Summary of the parameters derived from the modeling with 0.1~\micron\ 
and 0.5~\micron\ grains for the observations at two epochs.}

\begin{tabular}{l c c}\hline
Parameter & 2015 July 08  & 2016 March 23  \\
          & phase 0.92    & phase 0.54  \\
\hline
$T_{\rm eff}$ (K) & 2500 &  2140 \\
$L_{\star}$ (\LSOL) & 5130 &  3200 \\
Inner radius    & 1.9  &  1.3  \\
Outer radius    & 3.0  &  10.0 \\
Density exponent & 3   &  3 \\ 
\hline
\tauV\ (0.1~\micron) & $\le 0.2$  & 0.8 \\
$M_{\rm dust}^{0.1\mu{\rm m}}$ (\MSOL) & $\le 8.8\times10^{-10}$  & $4.9 \times 10^{-9}$ \\
\hline
\tauV\ (0.5~\micron) & 0.1  & $\le 0.1$ \\
$M_{\rm dust}^{0.5\mu{\rm m}}$ (\MSOL) & $3.9 \times 10^{-10}$ & $\le 5.5 \times 10^{-10}$ \\
\hline
$M_{\rm dust}^{0.5\mu{\rm m}}$/$M_{\rm dust}^{0.1\mu{\rm m}}$ & $\ge 0.42$ &
$\le 0.11$ \\
\hline
\label{model_res}
\vspace*{-7mm}

\end{tabular}
\end{center}
\end{table}

\section{Discussion}
\label{sect_discuss}

Our modeling of the second-epoch SPHERE-ZIMPOL data suggests the predominance 
of 0.1~\micron\ grains, noticeably smaller than the 0.5~\micron\ found 
for the first-epoch data. 
One may wonder if the presence of small, 0.1~\micron\ grains was not found 
by the modeling in Paper~I, simply because we assumed a single grain size. 
Therefore, we computed models with 0.1~\micron\ and 0.5~\micron\ 
\alumina\ grains and compared with the 2015 data in the same manner as 
described in Paper~I. 
We compared the models with the \IP\ maps at 645, 717, and 820~nm normalized 
with its peak value and the fraction of scattered light in the 
near-IR derived by Norris et al. (\cite{norris12}). 
In the modeling of the 2015 data, the $I$ and \PL\ maps 
were not used because the PSF references taken with Strehl ratios much lower 
than those for \whya\ made it impossible to convolve the model images 
with PSFs appropriate for comparison with the observed data. 
The parameters of the central star were set to be identical with those 
adopted in Paper~I.  We set the inner and outer boundary radii as well as 
the density distribution exponent to be the same for 0.1~\micron\ and 
0.5~\micron\ grains, adopting the values derived in Paper~I: 
the inner and outer radius of 1.9 and 3.0~\RSTAR, respectively, and the 
density distribution exponent of 3. 
We found out that the models with $ \tauV\ \!\! \le 0.2$ (0.1~\micron\ grains) 
and \tauV\ = 0.1 (0.5~\micron\ grains) can reproduce the \IP\ maps obtained 
at the first epoch and the fraction of scattered light in the near-IR.  
The corresponding dust mass is $8.8\times10^{-10}$~\MSOL\ and 
$3.9\times10^{-10}$~\MSOL\ 
for the 0.1 and 0.5~\micron\ grains, respectively. 
Therefore, it is possible that grains as small as 0.1~\micron\ 
were present in 2015. 

The parameters of the 0.1~\micron +0.5~\micron\ grain models derived 
for the 2015 and 2016 observations are listed in 
Table~\ref{model_res}.  It should be noted that the fraction of the large 
0.5~\micron\ grains is different at the two epochs. 
The ratio of the mass of the 0.5~\micron\ grains to that of the 0.1~\micron\ 
grains ($M_{\rm dust}^{0.5\mu{\rm m}}$/$M_{\rm dust}^{0.1\mu{\rm m}}$) 
is higher than 0.42 at pre-maximum light in 2015, while it is lower than 0.11 
at minimum light in 2016. Therefore, our modeling at two epochs suggests 
the predominance of grains as large as 0.5~\micron\ at pre-maximum light 
and the predominance of 0.1~\micron\ grains at minimum light.

Gobrecht et al. (\cite{gobrecht16}) present non-equilibrium gas-dust 
chemistry models for the Mira star IK~Tau.  Given its larger variability 
amplitude and its higher mass-loss rate than that of \whya, IK~Tau is 
considered to be more evolved than \whya. 
Their models do not include the dynamical effects caused by the 
radiation pressure on dust grains. Still, it may be worth comparing 
the results of such theoretical models with the phase dependence of 
the grain size suggested by our modeling.  Their models show that the 
passage of shocks destroys grains completely, and small grains start 
to form only half a period after the passage of the shocks, followed by 
efficient grain growth. 
The change in the characteristic grain size between the two epochs can be 
interpreted based on this theoretical prediction. 
Possibly, the 
first epoch at pre-maximum light in 2015 may correspond to the phase 
when the efficient grain growth was occurring, while the second epoch 
at minimum light may correspond to the phase when small grains just started to 
form. 
Their models also 
show that silicate dust forms at radii larger than 3.5~\RSTAR, while the 
formation of \alumina\ dust is restricted at radii smaller than 2~\RSTAR. 
It is possible that the polarization detected over the more extended region 
in 2016 may be attributed to the formation of silicate.

In contrast to the models of Gobrecht et al. (\cite{gobrecht16}), 
H\"ofner et al. (\cite{hoefner16}) present dynamical models in which 
the grain growth and the radiation pressure on dust grains are taken into 
account in a self-consistent manner, 
but the amount of the seed nuclei for dust formation is treated 
as a free parameter. 
They also included the formation of silicate mantle on \corundum\ 
grains, which was previously proposed by, e.g., 
Kozasa \& Sogawa (\cite{kozasa97a}, \cite{kozasa97b}). 
The models show the pulsation phase-dependence of the grain size 
and the formation of silicate mantle on \corundum\ core farther out 
than \corundum\ grains (see their Fig.~9). 
Therefore, the formation of grains in the extended region found in 2016 
may indicate the formation of silicate mantle on \corundum\ grains, 
instead of pure silicate as modeled by Gobrecht et al. (\cite{gobrecht16}). 
However, given that we cannot differentiate between 
the models with \corundum, 
\forsterite, and \enstatite\ as mentioned in Sect.~\ref{sect_modeling}, 
it is not possible to draw a definitive conclusion about the nature of 
the dust found in the extended region. 
The models of H\"ofner et al. (\cite{hoefner16}) also suggest that 
the radiation pressure on 
\corundum\ grains forming very close to the star is too low to drive 
mass loss, and \corundum\ grains with silicate mantle forming farther out 
may be more efficient in driving stellar winds by the scattering of stellar 
photons. 
SPHERE-ZIMPOL observations with higher cadence will help us to better 
understand the dynamics of the dust clouds and clarify where the effective 
wind acceleration takes place.

\section{Concluding remarks}
\label{sect_concl}

We have presented second-epoch visible polarimetric imaging observations 
of the AGB star \whya\ with VLT/SPHERE-ZIMPOL at minimum light (phase 0.54), 
which is distinct from the first-epoch observations at pre-maximum light 
(phase 0.92).  We have also reported on high-spectral resolution 
long-baseline interferometric observations of \whya\ 
near 2.3~\micron\ with VLTI/AMBER near minimum light.  

The polarized intensity maps obtained at five wavelengths from 645 to 820~nm 
with SPHERE-ZIMPOL reveal three clumpy dust clouds at 34--50~mas 
(1.4--2~\RSTAR).  Comparison of the data obtained at two epochs (phase 0.92 
and 0.54) shows clear time variations in the clumpy dust clouds close to 
the star. 
The degree of linear polarization measured at minimum light is 
systematically higher than that measured at pre-maximum light, and 
the polarization signals were detected in a noticeably more extended region 
compared to the first epoch. 
The continuum-subtracted \Ha\ emission observed at minimum light extends 
to $\sim$150~mas, 
also exhibiting time variations between pre-maximum light and minimum light. 
The continuum-subtracted TiO image obtained at minimum light 
reveals TiO emission extending to $\sim$150~mas ($\sim$6~\RSTAR), overlapping 
with the \Ha\ emission. 

Our VLTI/AMBER data have also revealed time variations in the extended 
atmosphere. The fitting to the visibilities at the continuum wavelengths 
near 2.3~\micron\ 
with a power-law-type limb-darkened disk has resulted in a limb-darkened 
disk diameter of $49.1 \pm 1.5$~mas and a limb-darkening parameter of 
$1.16 \pm 0.49$, which means that the atmosphere is more extended with 
weaker limb-darkening compared to pre-maximum light. 

Our 2-D Monte-Carlo radiative transfer modeling shows that the 
SPHERE-ZIMPOL data obtained at minimum light 
can be explained by a shell with $\sim$0.1~\micron\ grains 
of \corundum\ or \forsterite\ or \enstatite\ with a 550~nm optical depth of 
$0.6 \pm 0.2$ and an inner and outer radius 
of 1.3~\RSTAR\ and $10 \pm 2$~\RSTAR, respectively. 
Our modeling suggests the predominance of 0.1~\micron\ grains, in marked 
contrast to the predominance of large ($\sim$0.5~\micron) grains at 
pre-maximum light. 
This implies that small grains might just have started to form at 
minimum light, while the efficient grain growth might already have been 
taking place at pre-maximum light. 

High-cadence monitoring of the time evolution of the clumpy dust 
clouds as well as the extended \Ha\ and TiO emission 
is of great important for understanding the interplay between 
the pulsation-induced shocks and the molecule and dust formation. 
Given that the morphology of the images of R~Dor changed within 48~days 
(Khouri et al. \cite{khouri16}), SPHERE-ZIMPOL observations of \whya\ 
as frequent as every few weeks are useful for clarifying the physical 
process behind the time variations. 
Milliarcsecond-resolution velocity-resolved aperture-synthesis imaging of the 
extended atmosphere with VLTI/AMBER or VLTI/GRAVITY 
(Eisenhauer et al. \cite{eisenhauer08}) is crucial for examining whether or 
not the gas at the locations of the dust clouds shows systematic outward 
motions. 
Moreover, mid-IR spectro-interferometric imaging with VLTI/MATISSE 
(Lopez et al. \cite{lopez14}) will allow us to extract spatially resolved 
dust spectra, which will be indispensable for better understanding 
dust chemistry within $\sim$10~\RSTAR.

\begin{acknowledgement}
We thank the ESO Paranal team for supporting our SPHERE and AMBER 
observations. 
We are also grateful to the referee Susanne H\"ofner for her valuable 
comments, particularly on the interpretation of the extended \Ha\ emission. 
This research made use of the \mbox{SIMBAD} database, 
operated at the CDS, Strasbourg, France.  
We acknowledge with thanks the variable star observations from the AAVSO 
International Database contributed by observers worldwide and used in 
this research.
K.O. acknowledges a grant from Universidad Cat\'olical del Norte. 
\end{acknowledgement}

\end{document}